\documentclass{aa}
\usepackage{graphics,psfig}

\begin{document}

%
   \title{The Dynamics of the Circumnuclear Disk and its environment in the Galactic Centre}

   \author{B.~Vollmer\inst{1} \and  W.J.~Duschl\inst{2,1}}

   \offprints{B.~Vollmer, e-mail: bvollmer@mpifr-bonn.mpg.de}

   \institute{Max-Planck-Institut f\"ur Radioastronomie, Auf dem H\"ugel 69,
          D-53121 Bonn, Germany. \and
          Institut f\"ur Theoretische
          Astrophysik der Universit\"at Heidelberg, Tiergartenstra{\ss}e 15,
              D-69121 Heidelberg, Germany.}

   \date{Received / Accepted}

   \titlerunning{Dynamics of the CND and its environment}

\abstract{
We address the question of the dynamics in the inner 50~pc of the
Galactic Centre. In a first step we investigate the cloud--cloud collision
rate in the Circumnuclear Disk (CND) with the help of a three
dimensional N--body code using gas particles that can have inelastic
collisions. The CND might be a longer lived structure than previously assumed.
The whole disk--like structure of the CND can thus survive for several
million years. A realistic simulation of the CND shows the observed disk
height structure. In a second step the environment of the CND
is taken into account.
Retrograde and prograde encounters of a cloud of several
10$^{4}$~M$_{\odot}$ falling onto an already existing nuclear disk
using different energy loss rates per collision are simulated.
The influence of the energy loss rate per collision on the evolution
of the mass accretion and cloud collision rates is strongest
for a prograde encounter.
A composite data cube of two different snapshots of a prograde encounter
together with the CND shows striking similarity with the observed Sgr~A cloud
complex. The current appearance of the
Galactic Centre environment can thus be explained by at least two dynamically
distinct features together with the CND.
The current mass accretion rate within the CND ranges between
10$^{-3}$ and 10$^{-4}$~M$_{\odot}$\,yr$^{-1}$.
It can rise up to several 10$^{-2}$~M$_{\odot}$\,yr$^{-1}$
during massive accretion events.
}

\maketitle

\section{Introduction}

The Galactic Centre region is an ideal laboratory for studying the fueling
of a central black hole in great detail. Eckart \& Genzel (1996)
have observed the proper motions of bright stars in the central few
arcseconds around the non-thermal radio continuum source Sgr~A$^{*}$.
They concluded that there is strong evidence for a
$\sim$2.5\,10$^{6}$~M$_{\odot}$
central dark mass located within $\leq$0.015~pc of Sgr~A$^{*}$ with a
mass density of at least 6.5\,10$^{9}$~M$_{\odot}$\,pc$^{-3}$.
This high density excludes the possibility that the central mass concentration
is a central star cluster. Sgr~A$^{*}$ is surrounded by a huge H{\sc ii} region
Sgr~A West with a size\footnote{We assume 8.5 kpc for
the distance to the Galactic Centre} of 2.1$\times$2.9 pc, which was first
observed by Ekers et al. (1975). The ionized gas in this region forms a spiral
pattern (see e.g. Lo \& Claussen 1983, Lacy et al. 1991) and is therefore
called the {\it Minispiral}. The western side of this feature represents
the inner edge of a large ring of molecular and atomic gas, the {\it
Circumnuclear Disk (CND)} which extends from $\sim$2 to $\sim$7 pc from the
centre. It was observed by several authors:
Gatley et al. (1986) (H$_{2}$), Serabyn et al. (1986) (CO,CS), G\"{u}sten
et al. (1987) (HCN), DePoy et al. (1989) (H$_{2}$), Sutton et al. (1990)
(CO), Jackson et al. (1993) (O{\sc i}, HCN), Marr et al. (1993) (HCN),
Coil \& Ho (1999, 2000) (NH$_{3}$), and Wright et al. (2001) (HCN).
The main results of their investigations are:
\begin{itemize}
\item
The CND has a hydrogen mass of a few 10$^{4}$~M$_{\odot}$,
\item
the disk is very clumpy with an estimated volume filling factor of
$\Phi_{\rm V} \sim 0.01$,
\item
the clumps have masses of $\sim$30~M$_{\odot}$, sizes of $\sim$0.1~pc, and
temperatures $\geq$100~K.
\end{itemize}
It is now clear that the CND is interacting with the surrounding molecular
clouds, but it is still a matter of debate where these connections
are located (Zylka et al. 1990, Coil \& Ho 1999 2000, Wright et al. 2001).

Krolik \& Begelman (1988) constructed a clumpy disk model for AGNs where
cloud--cloud collisions are responsible for the energy and momentum
transport and thus for the viscosity. Shlosman et al. (1990) discussed
the possibility of fueling an AGN by the means of a cloudy disk.
Sanders (1998) investigated the accretion process in the
Galactic Centre on the basis of `sticky particle' calculations.
He concluded that the gas features observed within 10~pc from
the Galactic Centre can be understood in terms of tidal capture and
disruption of gas clouds on low angular momentum orbits in a potential
containing a point mass. The inner edge of the CND is explained
by the formation of a `dispersion ring', an asymmetric elliptical torus
precessing counter to the direction of rotation. This feature can be
maintained for many orbital periods.
These calculations follow the orbital motions of 4000 particles, whereas
the number of  observed clouds in the CND is $\sim$500.
Less clouds means less collisions and thus a lower viscosity (see e.g.
Pringle 1981). This has influences on the way the dispersion ring forms.

In a previous article, we constructed a self-consistent model to describe
the physical and kinematical state of the CND (Vollmer \& Duschl 2001a).
The gas clouds were described as isothermal spheres embedded in an
H{\sc ii} region. The disk structure formed by the clouds was described by
a quasi standard  continuous accretion disk using adequately averaged
parameters of the discrete cloud model. We succeeded in reproducing
observed quantities as the cloud mass, the cloud radius, the electron
density at the cloud edge, the electron density of the H{\sc ii} region,
the emission measure, the disk height, and the dispersion velocity.
A major result was that the collisional time scale for one cloud
is several Myr. The isolated CND might thus be much longer lived than
previously assumed (G\"{u}sten et al. 1987).

In Vollmer \& Duschl (2001b)  we pointed out a possibility to create the inner edge
of the CND: as the clouds approach the Galactic Centre only those
with densities high enough to resist tidal disruption can survive.
At a certain distance from the Galactic Centre, the clouds with high enough
densities to survive become too heavy and will fragment and/or collapse.
This critical distance depends on the radiation field because it
determines the radius of the clouds. Assuming realistic conditions
for the Galactic Centre, we ended up with an inner edge at $\sim$2~pc.

The present article is devoted to the dynamics of the gas clouds in the CND.
With the help of an improved  numerical 3D collisional cloud model we
investigate the collisional time scales of the CND, which is assumed
not to interact with the surrounding molecular clouds.
A more realistic way to model the Galactic Centre region is to
include the molecular clouds, whose projected distances to the CND are small
(see e.g. Mezger et al. 1989, Zylka et al. 1990). In the second part
of this article we take this environment into account in our model
and deduce more realistic collision rates for a gas clouds
falling into the Galactic Centre with an already existing CND.
A possible scenario for the star formation history in the
Galactic Centre is presented.

\section{The model \label{sec:model}}

The numerical simulations must take into account that the CND consists
of several hundred gas clouds. The present code treats
each cloud as a point with a given mass and radius.
The mass--radius relation for the clouds might be given by the virial theorem
\begin{equation}
r_{\rm cl}=0.478\times M_{\rm cl}[{\rm M_{\odot}}]/T_{\rm cl}[{\rm K}]\ {\rm pc}
\end{equation}
where $r_{\rm cl}$ is the cloud radius, $M_{\rm cl}$ the cloud mass, and
$T_{\rm cl}$ the cloud temperature.
The temperature distribution with respect to the distance from the
Galactic Centre is assumed to be
\begin{equation}
T_{\rm cl}(R)=224/\sqrt{R[{\rm pc}]}\ {\rm K}
\label{eq:temp}
\end{equation}
where $R$ is the distance of the cloud from the Galactic Centre.
This distribution is consistent with the observed UV radiation field
(Vollmer \& Duschl 2001a) and the deduced gas temperature (see e.g.
G\"usten et al. 1987, Sutton et al. 1990).
On the other hand, Vollmer \& Duschl (2001a) have shown that the clouds
in the CND have a constant radius $r_{\rm cl} \simeq 0.05$~pc if one takes
the ionization front due to the external UV radiation field into account.
In the present study we use both mass--radius relations.

We follow the orbits of these clouds in the three dimensional gravitational
potential. The radial distribution of the total enclosed mass $M(R)$ is given by
\begin{equation}
M(R)=M + M_{0}R^{\frac{5}{4}}
\label{eq:gravpot}
\end{equation}
where $M=3\,10^{6}$~M$_{\odot}$ is the mass of the central black hole,
and $M_{0}=1.6\,10^{6}$~M$_{\odot}$/pc$^{\frac{5}{4}}$ describes the mass
distribution of the stellar content. This is close to the
findings of Eckart \& Genzel (1996).
When orbiting around the Galactic Centre, the clouds can have
inelastic collisions. For the search of the next neighbour of a cloud a
Barnes \& Hut (1986) treecode is used.
During these collisions clouds can exchange mass
and larger clouds can grow through coalescence. The result of
a collision can be one (coalescence), two (mass exchange), or many
fragments (fragmentation). In our model, we limit the fragmentation case to
a maximum of 3 fragments.
Let the radius of the first cloud be $r_{1}$, that of the second
cloud $r_{2}$. Let the impact parameter be $b$, the velocity of
the fragment $v_{\rm f}$, and the escape velocity $v_{\rm esc}$.
We follow the prescriptions of Wiegel (1994):
\begin{itemize}
\item
for $r_{1}-r_{2} < b < r_{1}+r_{2}$:\\ fragmentation
\item
for $b \le r_{1}-r_{2}$ and $v_{\rm esc} > v_{\rm f}$:\\
mass exchange
\item
for $b \le r_{1}-r_{2}$ and $v_{\rm esc} \le v_{\rm f}$:
\\ coalescence
\end{itemize}
In order to avoid the production of large number of too small clumps
($M<10$~M$_{\odot}$), we treat a fragmentary collision, which would give rise
to a cloud with $M<10$~M$_{\odot}$, as a mass--exchange collision in
adding the mass of the potential third cloud to that of the lightest
colliding clouds. This procedure ensures that the arising cloud
mass spectrum has a maximum between 10 and 15~M$_{\odot}$.

The integration of the ordinary differential equation is done with the
Burlisch-Stoer method (Stoer \& Burlisch 1980) using a Richardson
extrapolation and Stoermer's rule.
This method advances a vector of dependent variables $y(x)$ from a point
$x$ to a point $x+H$ by a sequence of $n$ substeps.
Thus, the initial timestep $H$ is divided subsequently into
$n$=2, 3, 4, etc. substeps. At the end the solution
of $y(x+H)$ is extrapolated and an error can be estimated.
The size of the timestep is adaptive and linked to the estimated
error of the extrapolation. This error is normalized by the
values of the distance covered during the last timestep and the velocity
of each particle. The error level for acceptance of the extrapolated
solution is a free parameter and has to be adapted to the
physical problem treated. For a relative error level
$\Delta r_{i} / r_{\rm norm} = \Delta v_{i} / v_{\rm norm} < 0.1$,
where $i$=1, 2, 3, $r_{\rm norm}=10^{-2}$~pc and
$v_{\rm norm}=5\,10^{-2}$~km\,s$^{-1}$
the Courant criterion is fulfilled for each timestep.

The collisions are evaluated at each timestep $h=H/n$ and only those,
which appear for all sequences $n$, are taken into account.
Therefore, the relative error level $\epsilon$ is crucial for the
value of the obtained collision rates. We adopt the strategy to
chose the error level in a way to match the theoretical collision
rates.

\section{The normalization of the collision rate \label{sec:normalization}}

With the fixed precision of the calculations it is possible to
compare the model collision time for one cloud with the theoretical one:
\begin{equation}
t_{\rm coll} \simeq (n_{\rm cl}\,\sigma_{\rm cl}\,v_{\rm cl})^{-1}
\end{equation}
where $n_{\rm cl}$ is the local cloud density, $\sigma_{\rm cl}
=\pi\,r_{\rm cl}^{2}$ the cross section of the cloud, and $v_{\rm cl}$
the cloud velocity. The comparison of the mean collision time
for one cloud are made with the simplest spatial configuration,
namely a spherical volume. The clouds are initially distributed
uniformly in a sphere with a radius of 4~pc. The initial velocities
are uniformly distributed between 20~km\,s$^{-1}$ and 40~km\,s$^{-1}$.
The gravitational potential is due to the following mass
distribution: $M(R)=M_{0}\,R^{\frac{5}{4}}$ with $M_{0}$ from
Eq.~(\ref{eq:gravpot}). In order to check our numerics, we made one set
of simulations where the potential collisions are counted, but not executed.
In this case there is only negligible angular momentum transfer due to
gravitational interactions between the clouds and the spatial
distribution of the clouds stays approximately constant during the simulation.

The influence of the mass--radius relation on the collision rate is tested
with two different relations:
\begin{itemize}
\item
RUN1: $r_{\rm cl} \propto M_{\rm cl}\,\sqrt{R}$.
In this case the cloud temperature adjusts to the
temperature given in Eq.~(\ref{eq:temp}). Since there is
cloud heating due to cloud--cloud collisions, it is assumed that the
dynamical time scale is much larger than the cooling time scale
due to radiation.
\item
RUN2: $r_{\rm cl} = {\rm const.}$.
In this case an external UV radiation field is assumed
that creates an ionization front on the cloud surface which is
directed to the Galactic Centre. The location of this ionization front
does not depend on the clouds' distance to the Galactic Centre
(Vollmer \& Duschl 2001a), i.e. the cloud radius is constant.
\end{itemize}

\begin{figure}
    \resizebox{8cm}{6cm}{\includegraphics{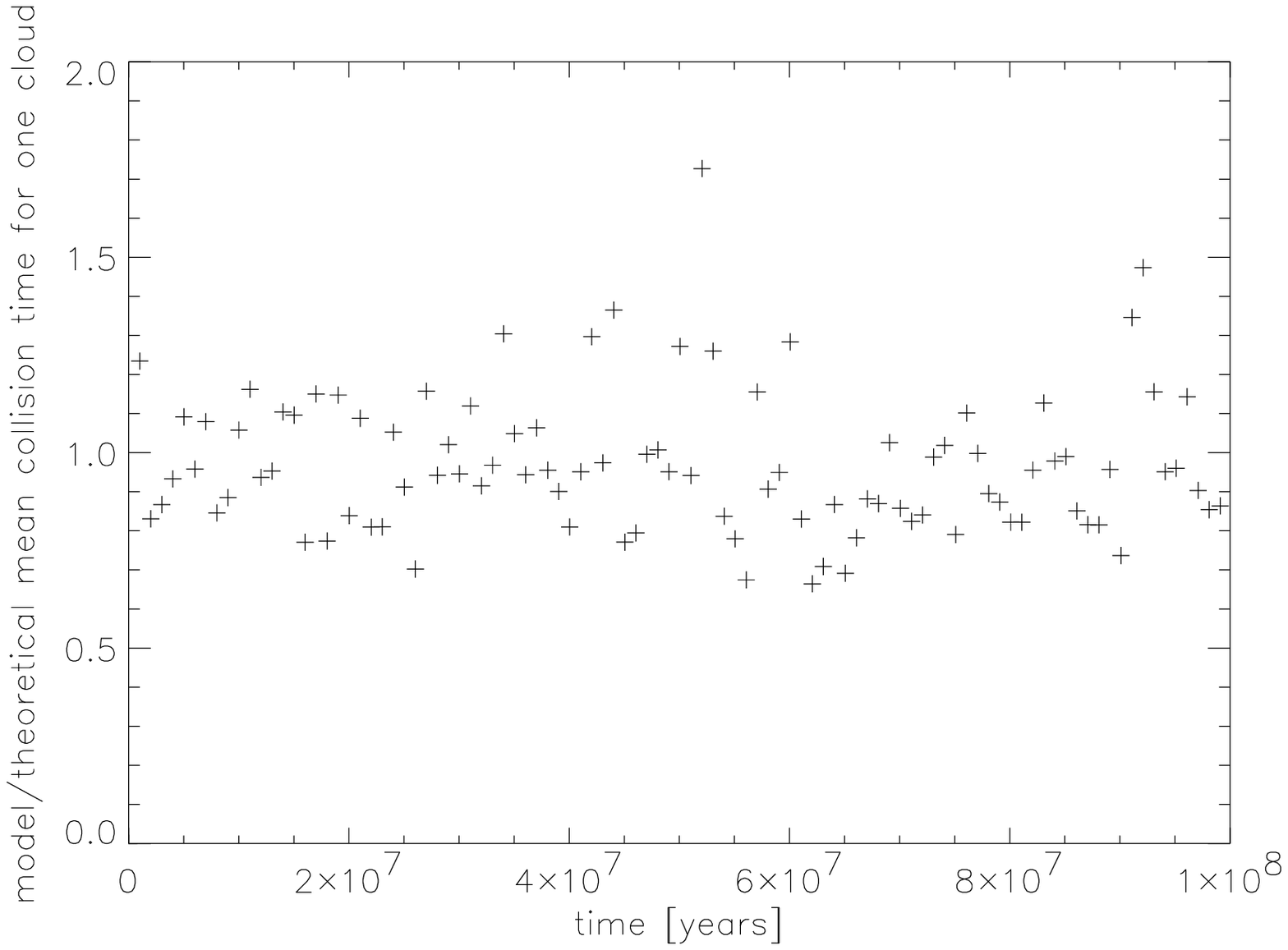}}
    \resizebox{8cm}{6cm}{\includegraphics{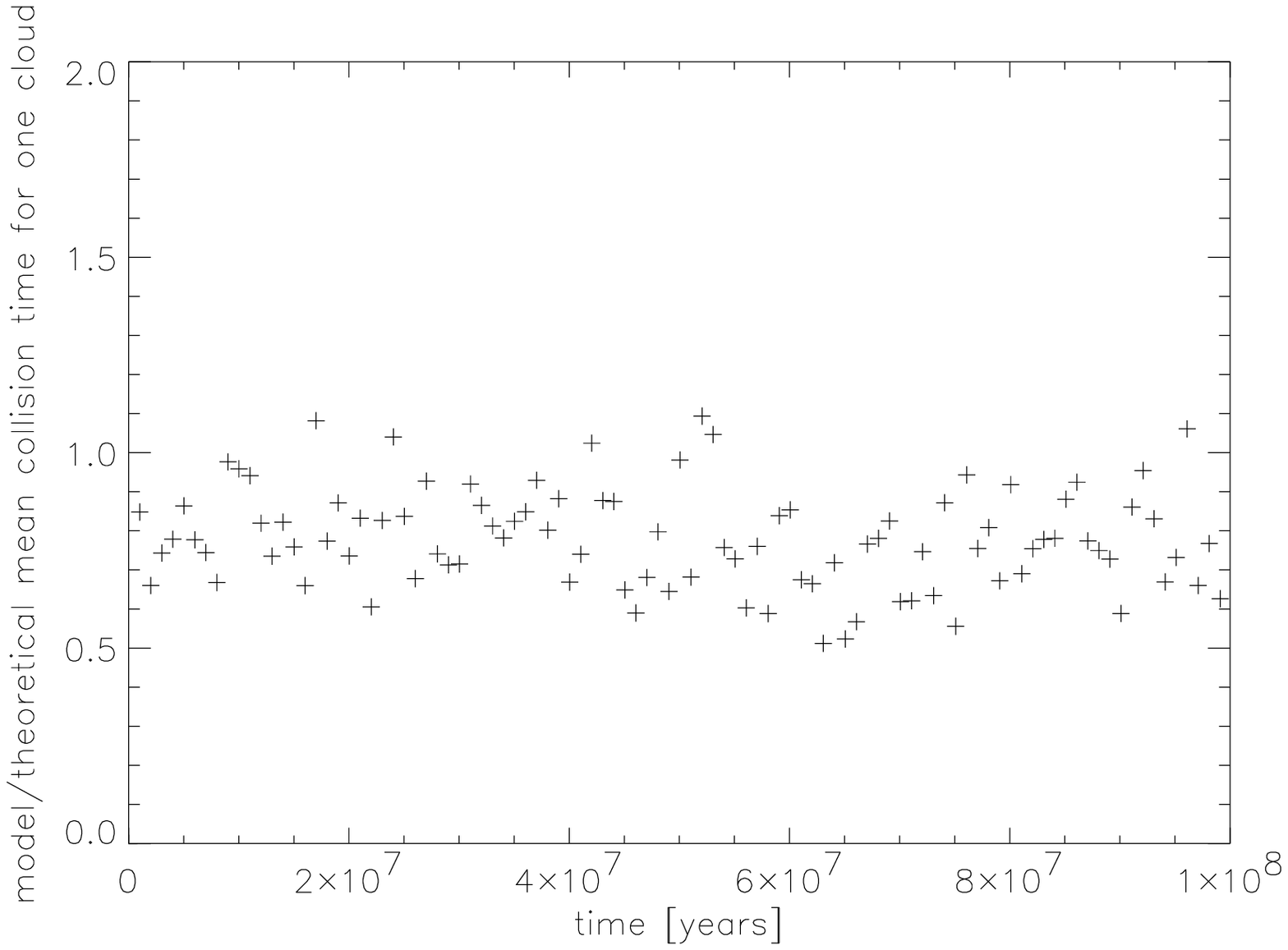}}
      \caption{ The ratio between the model and theoretical mean collision time
        without executed collisions. Left panel: RUN1. Right panel: RUN2.
        \label{fig:model_theo_withoutcoll_temp}
}
\end{figure}

The ratio of the model mean collision time of one cloud and the
theoretical one for RUN1 and RUN2 without executed collisions
is shown in Fig.~\ref{fig:model_theo_withoutcoll_temp}.
The theoretical mean collision time is given by
\begin{equation}
t^{\rm theo}_{\rm coll} \simeq \langle n_{\rm cl} \sigma_{\rm cl} v_{\rm cl}
\rangle ^{-1} \simeq \big(\langle n_{\rm cl} \rangle \langle \sigma_{\rm cl} \rangle
\langle v_{\rm cl} \rangle \big)^{-1}\ .
\label{eq:tcoll}
\end{equation}
The ratio for both simulations stays constant over 100~Myr. It is $\sim$1
for RUN1 and $\sim$0.75 for RUN2. Thus, both simulations reproduce the
theoretical collision time scale correctly within 25\%.
The difference between RUN1 and RUN2 lies in the density gradient of the cloud
distribution. Clouds at the outskirts
of the sphere are larger for RUN1. Therefore, the average theoretical collision
time scale is smaller, and the ratio between the model and the theoretical
collision time scales is larger. Since these clouds are located in a
region with a low particle density and do not contribute much
to the global model collision time scale, the ratio between the model and theoretical
collision time scale is larger.

\begin{figure}
    \resizebox{\hsize}{!}{\includegraphics{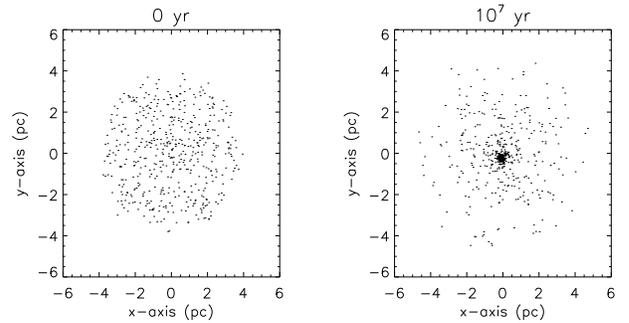}}
      \caption{Right: initial state of a simulation with collisions.
    Left: final state of the simulation after 10$^{7}$~yr.
    \label{fig:initial_sphere}
}
\end{figure}

\begin{figure}
    \resizebox{8cm}{6cm}{\includegraphics{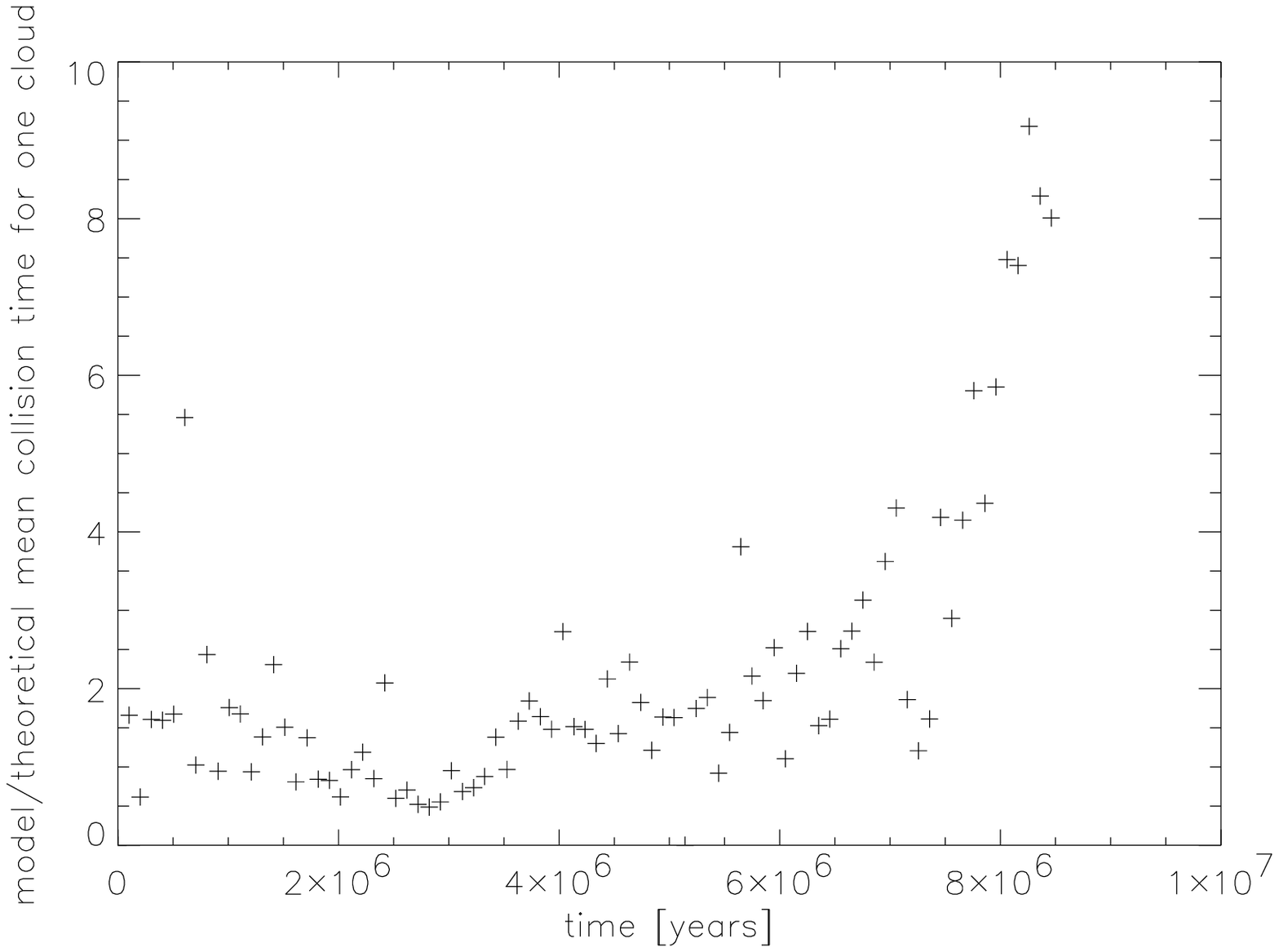}}
    \resizebox{8cm}{6cm}{\includegraphics{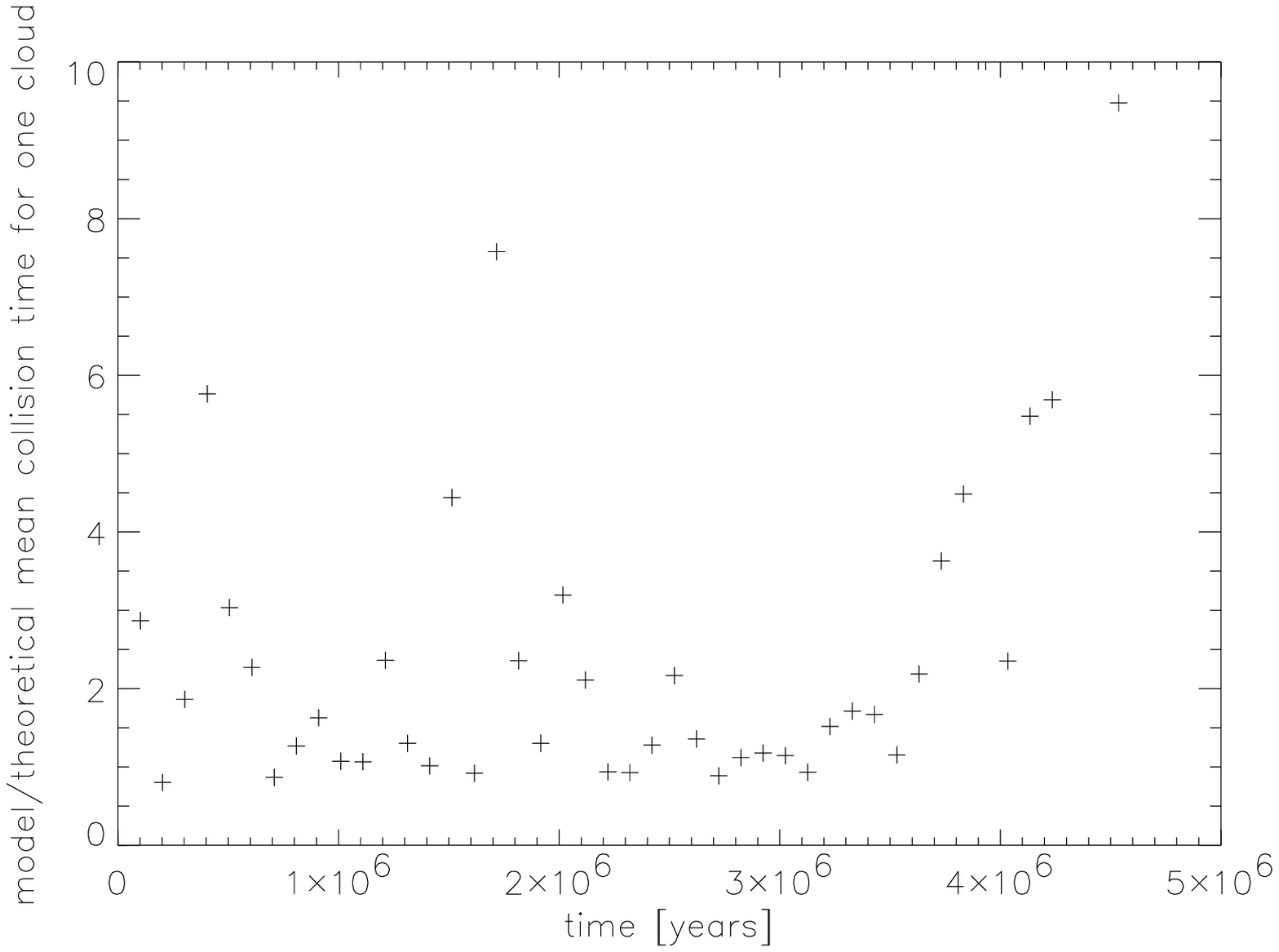}}
      \caption{ The ratio of the model mean collision time to the theoretical one
        with proceeding collisions. Left panel: RUN1. Right panel: RUN2.
        \label{fig:model_theo_withcoll_temp}}
\end{figure}

We then calculated another set of simulations with proceeding cloud--cloud
collisions. As the system evolves, the angular momentum transport
through these collisions, i.e. the viscosity, makes the particle density
increase in the centre. The initial and the final stage of such a simulation
is shown in Fig.~\ref{fig:initial_sphere}.
The ratio between the model and theoretical mean collision time
for RUN1 and RUN2 is shown in Fig.~\ref{fig:model_theo_withcoll_temp}.
The theoretical mean collision time is given by Eq.~(\ref{eq:tcoll}).

\begin{itemize}
\item
RUN1: During 6~Myr the ratio between the model and theoretical collision
time scales is $\sim$1.
As the spatial cloud distribution changes to higher central densities,
the number of collisions in the system is overestimated by Eq.~(\ref{eq:tcoll}),
i.e. the collisional time scale is underestimated. This makes the
ratio between the model and the theoretical collision time scale increase.
The critical cloud density for this increase is $n_{\rm cl} \sim 100$~clouds/pc$^{3}$.
\item
RUN2: We observe here the same trend as for RUN1. When the cloud distribution
contracts, the number of collisions is theoretically underestimated.
This effect also starts at a critical cloud density of
$n_{\rm cl} \sim 100$~clouds/pc$^{3}$.
\end{itemize}

We thus conclude that the mean collision time for one cloud,
which is resulting from the model, reflects the real value or
might underestimate it by $\sim$25\% if the cloud distribution
is not heavily concentrated.

\section{CND simulations as an isolated structure}

We have distributed 500 particles within a ring volume of constant height
$H$=1~pc and an inner radius of $R_{\rm in}$=2~pc and an outer
radius $R_{\rm out}$=7~pc. The initial velocities were Keplerian
using a gravitational potential made by the mass distribution described
in Eq.~(\ref{eq:gravpot}). A turbulent velocity of 20\% of the Keplerian
value was added in an arbitrary direction. The system evolved then
freely in the gravitational potential of a point mass
(the central black hole) and an extended mass distribution
(the central star cluster). Vollmer \& Duschl (2001b) have shown that
the clouds, which approach the Galactic Centre, become tidally disrupted
if they are not dense enough to resist. Those clouds with a high
enough central density to be stable at a radius of $R_{\rm crit} \sim$2~pc
will collapse and/or fragment. In any case, no cloud can survive
at radii closer than $R_{\rm crit}$. In our model we take this effect into
account by removing clouds with $R < R_{\rm crit}$.

\begin{figure}
    \resizebox{\hsize}{!}{\includegraphics{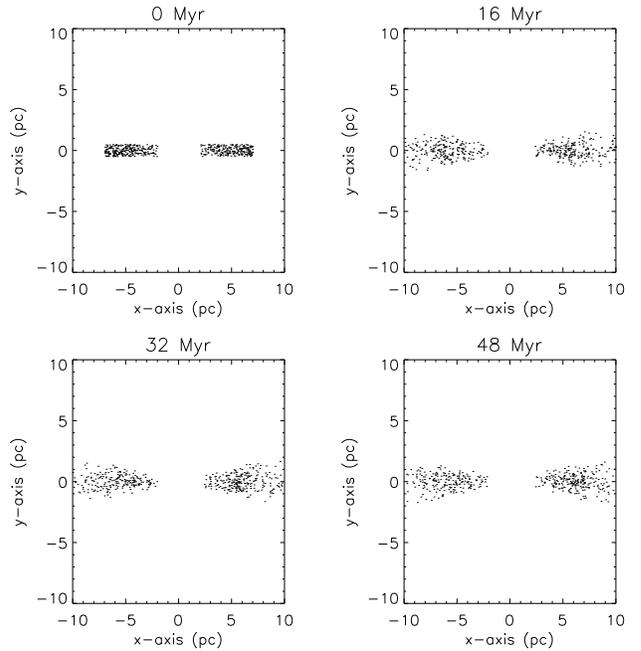}}
      \caption{The evolution of the cloudy disk model without collisions.
    The elapsed time is shown at the top of each frame. The $x$-axis shows
    the distance of the clouds to the centre, the $y$-axis
    shows the vertical distance with respect to the disk plane.
\label{fig:disk_evolution_withoutc}}
\end{figure}

\begin{figure}
\begin{center}
    \resizebox{8cm}{6cm}{\includegraphics{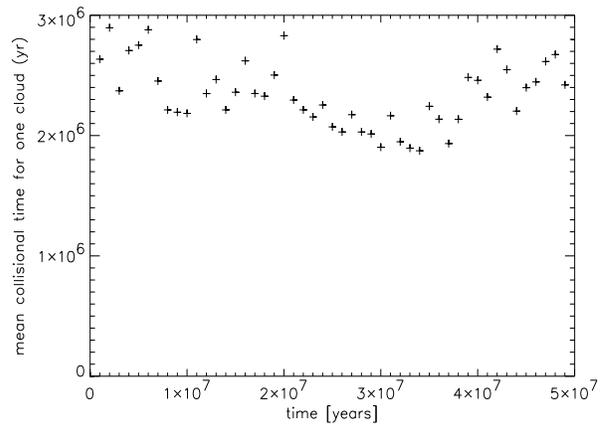}}
      \caption{The evolution of the collision time scale for one cloud
    for the simulation shown in Fig.~\ref{fig:disk_evolution_withoutc}.
\label{fig:mean_disk_withoutcoll_0.05}}
\end{center}
\end{figure}

As in Sect.~\ref{sec:normalization} we have made first a simulation
counting the collisions without executing them.
The cloud radius is $r_{\rm cl}=const=0.05$~pc.
Fig.~\ref{fig:disk_evolution_withoutc} shows different states of evolution of such
a disk-like structure.
The timesteps are $\Delta t = 16$~Myr. The first plot shows the
ring--like initial condition. After several Myr the disk reaches
a state of equilibrium. The disk thickness at a distance of 5~pc
is about 2~pc. The collisional time scale for one cloud is
shown in Fig.~\ref{fig:mean_disk_withoutcoll_0.05}.
It ranges between 2 and 3~Myr.

\begin{figure}
    \resizebox{\hsize}{!}{\includegraphics{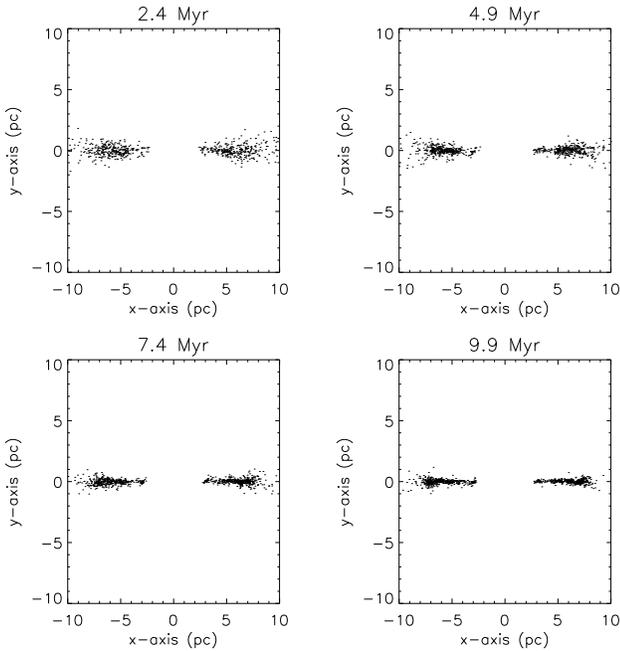}}
      \caption{The evolution of the cloudy disk model with collisions.
    The elapsed time is shown at the top of each frame. The $x$-axis shows
    the distance of the clouds to the centre, the $y$-axis
    shows the vertical distance with respect to the disk plane.
\label{fig:disk_evolution_withc}}
\end{figure}

In the next step we have made two simulations with collisions using the
two different mass--radius relations RUN1 and RUN2.
Since the evolution of the spatial cloud distribution is quasi identical
for RUN1 and RUN2, we show only that of RUN2 in Fig.~\ref{fig:disk_evolution_withc}.

\begin{figure}
    \resizebox{\hsize}{!}{\includegraphics{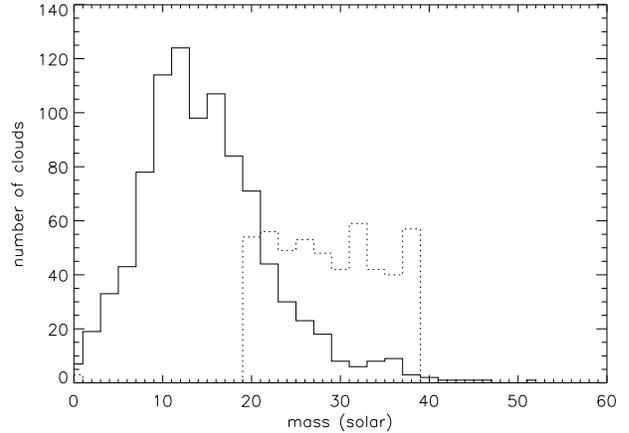}}
      \caption{The mass spectrum of the clouds. Dotted line: initial
    mass spectrum. Solid line: mass spectrum after 10~Myr.
\label{fig:massspec}}
\end{figure}

The observed thickness of the CND, i.e. 0.5~pc at a distance of 2~pc and 1.5~pc
at a distance of 5~pc from the Galactic Centre (G\"{u}sten et al. 1987) lies
between those of the model with and without collisions.
The initial mass distribution
18~M$_{\odot} < M_{\rm cl} < 38$~M$_{\odot}$ (Fig.~\ref{fig:massspec}
dotted line) smears out due to the inelastic collisions. The mass distribution
after 10~Myr is plotted as a solid line in Fig.~\ref{fig:massspec}.
The maximum has shifted to smaller cloud masses and a high mass tail
has been built. A part of these high mass clouds should have collapsed,
but this mechanism is not included in our code.

\begin{figure}
    \resizebox{8cm}{6cm}{\includegraphics{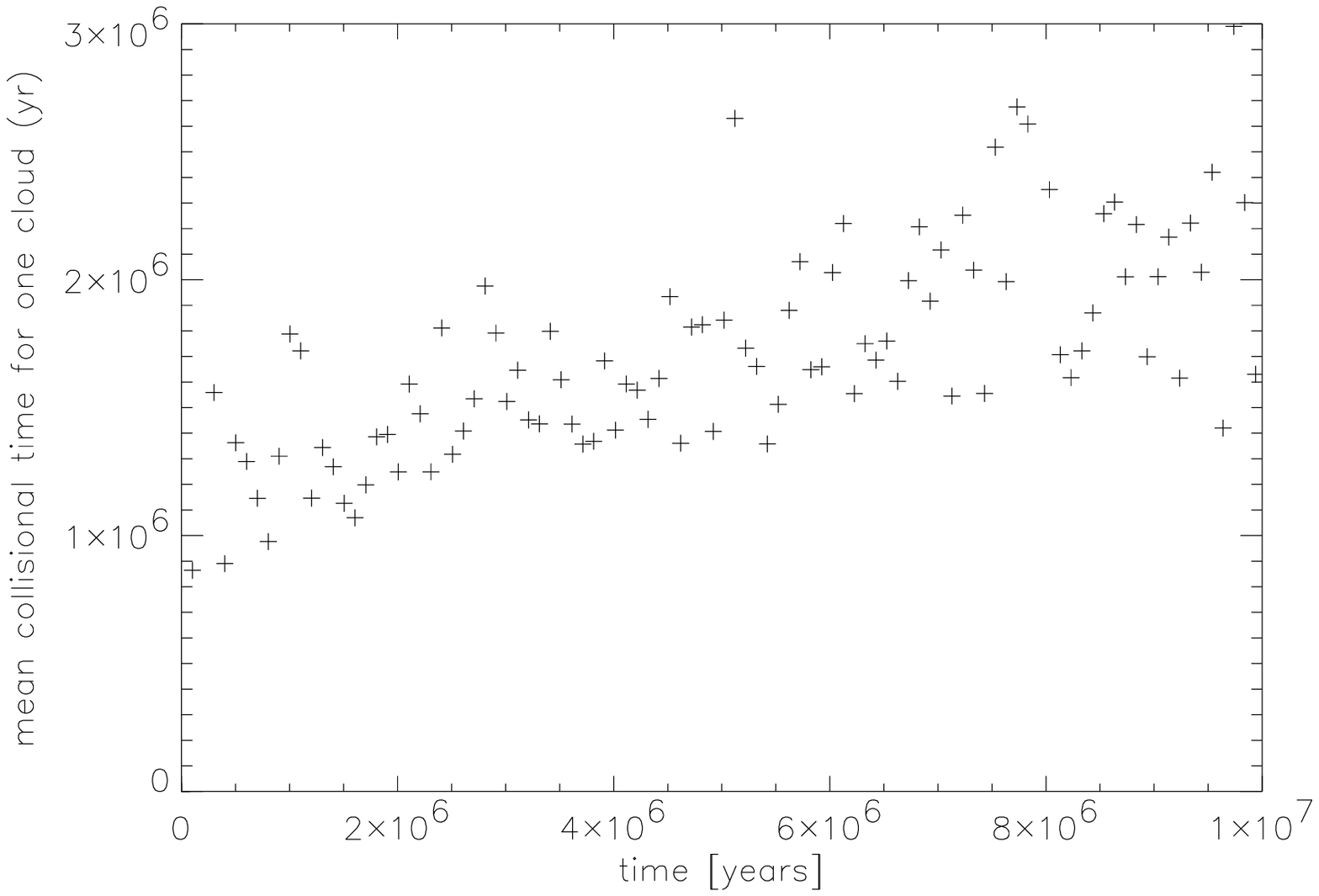}}
    \resizebox{8cm}{6cm}{\includegraphics{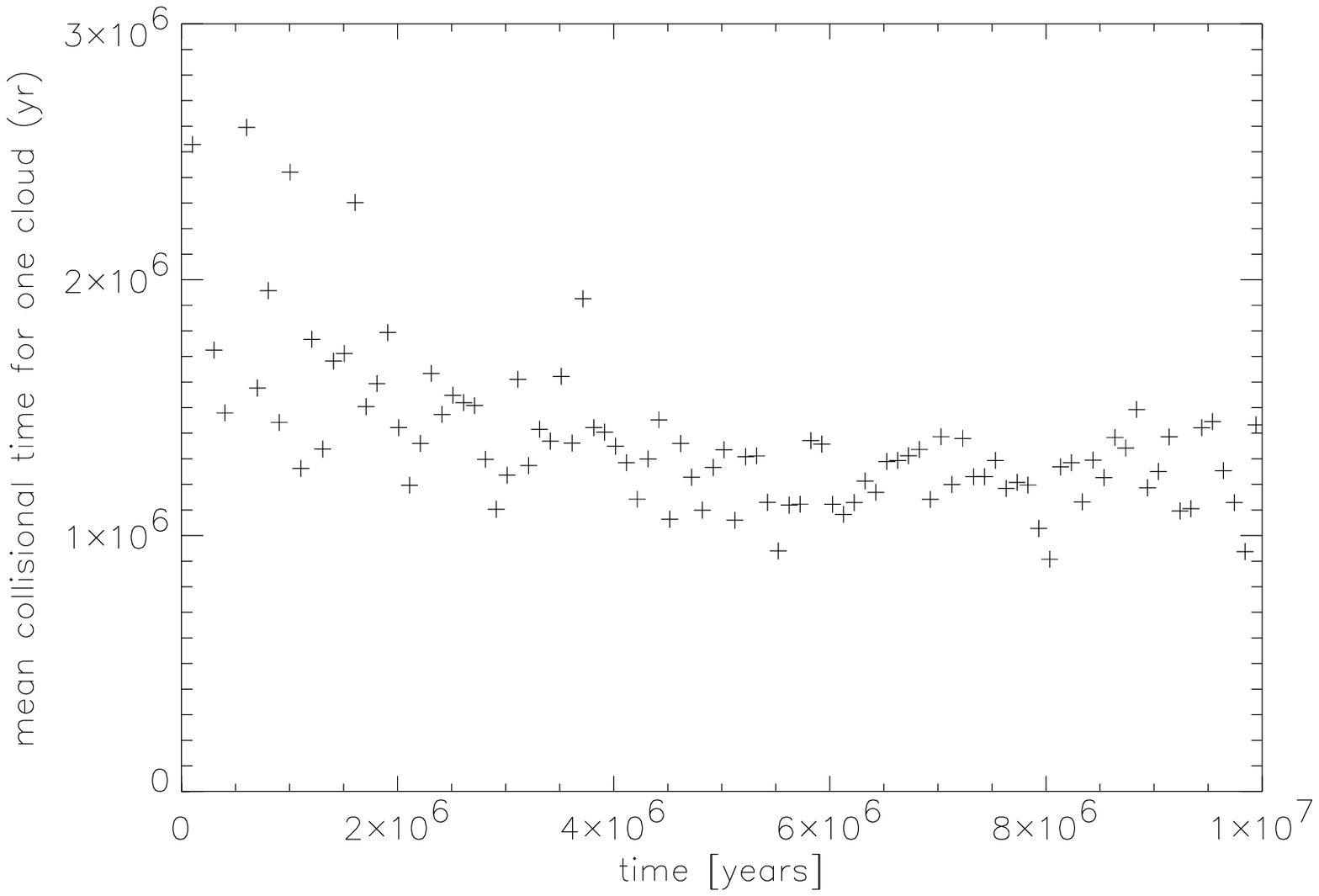}}
      \caption{Model mean collision times for the mass-radius relation
    Left panel: RUN1. Right panel: RUN2.
\label{fig:mean_disk_withcoll_temp}}
\end{figure}

The resulting evolution of the model mean collision time can be seen for
the mass--radius relation RUN1 and RUN2 in Fig.~\ref{fig:mean_disk_withcoll_temp}.
We observe an increasing mean collision time for the mass-radius relation
RUN1 (Fig.~\ref{fig:mean_disk_withcoll_temp} left panel).
This is, because the cloud distribution is shifted
to smaller masses. Thus the clouds are smaller and the collision time scale,
which is proportional to the square of the cloud radius, increases.
For the case of a constant radius independent of the cloud mass \
(RUN2; Fig.~\ref{fig:mean_disk_withcoll_temp} right panel)
the collision rate increases, because the number of clouds and thus
the cloud density increases.

Since the cloud density is always smaller than $n_{\rm crit}$ the results
for the mean collision time scales is robust. The derived mean collision time
scale for one cloud is of the order of 1--2~Myr.

In Vollmer \& Duschl (2001a) we gave a mean collision time scale of
$\sim$10~Myr. The factor 5 between both approaches can be understood
in the light of the analytical model for turbulent, clumpy accretion disks
(Vollmer \& Beckert 2002). In this model, the viscosity is given as
$\nu=Re^{-1}v_{\rm turb}l_{\rm driv}$, where $v_{\rm turb}$ is the
turbulent velocity dispersion and $l_{\rm driv}$ is the driving wavelength
for the turbulence. From the analytical model it follows that $l_{\rm driv}=H$,
where $H$ is the disk height. Thus one obtains $\nu=Re^{-1}v_{\rm turb}H$.
This viscosity prescription is also used in Vollmer \& Duschl (2001a).
The relation between the gas surface density $\Sigma$ and the mass accretion rate
$\dot{M}$ of the disk is (see, e.g., Pringle 1981)
\begin{equation}
3 \pi \nu \Sigma = \dot{M}\ .
\label{eq:pring}
\end{equation}
The collision time scale for one cloud is then $t_{\rm coll}=Re \Omega^{-1}$.
In Vollmer \& Duschl (2001a) we used $Re=1000$, whereas a Reynolds number
of $Re=20$ has to be used in the framework of Vollmer \& Beckert (2002),
which is consistent with our dynamical results (Fig.~\ref{fig:mean_disk_withcoll_temp}).
Vollmer \& Duschl (2001a) assumed thus a relatively large Reynolds number.
Consequently, they obtained a relatively large gas surface density (Eq.~\ref{eq:pring}),
i.e. a relatively large disk mass ($M_{\rm gas} \sim 1.5\,10^{5}$~M$_{\odot}$), \
about a factor of 10 larger than that what we have used here.

\section{Environmental effects \label{sec:env}}

In the previous Sections we have treated the CND as an isolated structure.
In reality this is not the case. Observation of molecular transitions (e.g.
CO: Sutton et al. 1990; CO, CS: Serabyn et al. 1986, HCN: Coil \& Ho 1999, 2000,
Wright et al. 2001) and mm observations
(e.g. Mezger et al. 1989, Dent et al. 1993) have shown that there
are several molecular cloud complexes in the vicinity of the
Galactic Centre. Sgr~A East Core, a compact giant molecular cloud with
a gas mass of several 10$^{5}$ M$_{\odot}$ and the giant molecular
clouds M-0.13-0.08 and M-0.02-0.07 form the Sgr~A cloud complex.
We will compare our simulations to the observational results of
Zylka et al. (1990), because their distinction of
different cloud complexes is made on kinematical grounds.
They obtained the following results:
M-0.13-0.08 has radial velocities in the range of $\sim$5--25km\,s$^{-1}$,
a total mass of $\sim$3\,10$^{5}$ M$_{\odot}$ and lies in front of the
Galactic Centre. M-0.02-0.07 consists of two different features.
(i) Sgr~East Core with a mass of $\sim$2\,10$^{5}$ M$_{\odot}$ and
(ii) a curved streamer of total mass $\sim$10$^{5}$ M$_{\odot}$ and
velocities between 25 and 65 km\,s$^{-1}$, which lies also in front
of the Galactic Centre. Thus, there is neutral gas with a mass
of several 10$^{5}$ M$_{\odot}$, which is presumably located
within the inner 50~pc around the Galactic Centre
(see, however, von Linden et al. 1993). In addition, there
are hints that they are kinematically connected to the CND.

\subsection{Initial conditions}

\begin{figure}
    \resizebox{\hsize}{!}{\includegraphics{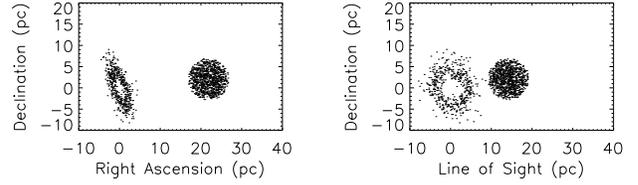}}
      \caption{Initial conditions for the cloud infall into the
    Galactic Centre. The axis correspond to: left: right ascension,
    declination; right: LOS ($z$-axis), declination.
    \label{fig:initcond}}
\end{figure}

We let evolve a model of the CND as described above with 500 particles and a
total mass of $\sim$1.5\,10$^{4}$~M$_{\odot}$ during 10~Myr.
We then added a spherical mass distribution of 1000 particles,
a total mass of 3\,10$^{4}$~M$_{\odot}$, and a size of 10~pc at a
distance of $\sim$30~pc. The initial values of the velocity
of these clouds were 30\% of the Keplerian value with an additional
velocity dispersion of 20\% of this value. Due to the sub-Keplerian
velocity, the cloud falls into the Galactic Centre colliding
with the CND. This initial configuration can be seen in Fig.~\ref{fig:initcond}.
We have made two different sets of simulations:\\
(i) a prograde encounter, i.e. the orbital angular momentum of the cloud
is close to that of the disk;\\
(ii) a retrograde encounter, i.e. the orbital angular momentum of the cloud
is opposite to that of the disk.
Technically, this is realized in mirroring  the $z$-axis for the disk clouds:
$z^{i}=-z^{i},\ v_{z}^{i}=-v_{z}^{i}$, where $i$ is the number of the cloud.

Furthermore, we varied the fraction of kinetic energy that is radiated
away during a cloud--cloud collision. This has important influences on the
collision and mass accretion rate of the system.

\subsection{The model evolution \label{sec:evolution}}

\subsubsection{Retrograde encounter}

\begin{figure}
    \resizebox{\hsize}{!}{\includegraphics{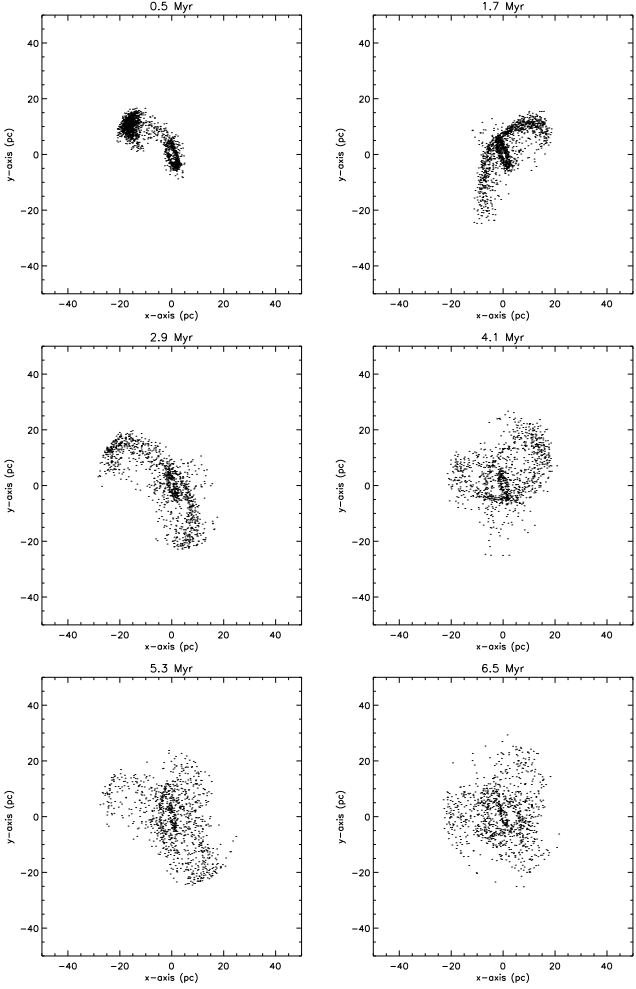}}
      \caption{The evolution of the cloud infall into the
    Galactic Centre as the observer would see it from the Earth.
    The elapsed time is plotted on the top of each frame.
    \label{fig:evolution}}
\end{figure}

The evolution of a system with no loss of kinetic energy per collision
can be seen in Fig.~\ref{fig:evolution}.
The timestep is $\Delta t$=1.2~Myr.
The cloud hits the CND for the first time at $t \simeq 0.3$~Myr.
The second collision takes place at $t \simeq 0.8$~Myr. The period
of the orbit is $t_{\rm orb}\sim$0.7~Myr. During its
orbiting the cloud is stretched due to the tidal shear.
The streamer becomes a filament of $\sim$40~pc within 1~Myr.
As the cloud collides for the first time with the CND, the latter
is heavily damaged and shows an azimuthally asymmetric cloud distribution.
Meanwhile, a counter rotating streamer is building up at the outer part of
the CND. After $\sim$3~Myr the CND becomes less and less prominent.
After $\sim$2~Myr the streamer crosses itself for the first
time and has a ``brezel'' form at $t \sim$3~Myr.
Later on, it begins to form a triangular feature. At the end of the
simulation ($t \sim$10~Myr), we observe this triangular feature
containing a ring--like rotating structure extending from $\sim$5 to
$\sim$10~pc and a counter rotating core (the former CND).
The simulation with 10\% energy loss per collision shows a more
pronounced ring--like structure and a less prominent rotating core.

\begin{figure}
    \resizebox{\hsize}{!}{\includegraphics{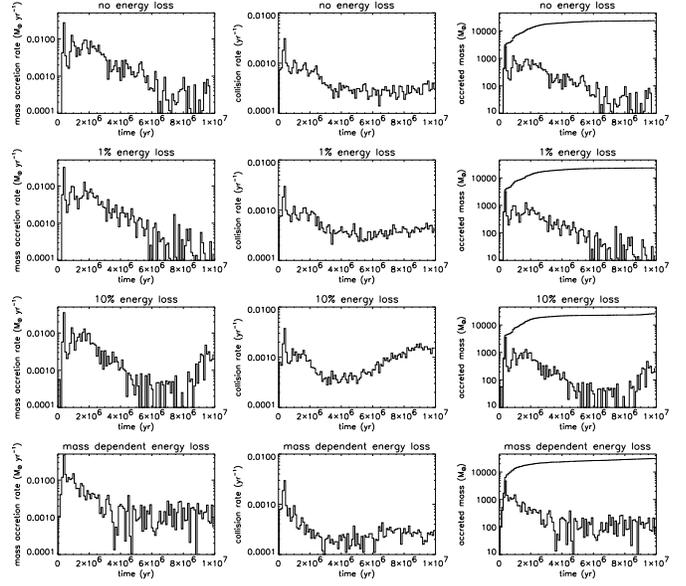}}
      \caption{1st column: mass accretion rate;
    2nd column: number of collisions per timestep $\Delta t$;
    3rd column: accreted mass per timestep $\Delta t$ and accumulate accreted mass.
    Rows: no energy loss, 1\%, 10\%, mass dependent loss of
    kinetic energy per collision.
    \label{fig:accrete}}
\end{figure}

This evolution can be also described by the number of collisions
and by the mass accretion rate. We define this mass accretion rate
by the accumulated mass of clouds per timestep $\Delta t$,
which have distances to the Galactic Centre less than 2~pc.
This evolution is shown in Fig.~\ref{fig:accrete} for four different
energy loss rates per collision: 0\%, 1\%, 10\%, and an energy loss
that is proportional to the fraction of the clouds' mass, which
participates in the collision (Krolik \& Begelman 1988).
The evolution of the mass accretion rate and the collision rate are
similar for all energy loss rates.
During the first collision at $t \sim$0.4~Myr the total accreted
mass is $\sim 3000~M_{\odot}$ within 0.1~Myr.
Nevertheless, this is not the major accretion event. This happens between 2 and 3~Myr,
when the tidally stretched cloud (streamer) is crossing itself.
The total accreted mass in this second event is $\sim$10$^{4}$~M$_{\odot}$
within 2~Myr. At later
stages of the simulation the mass accretion rate decreases steadily having
still some small peaks. The comparison with the collision rate
shows that the first accretion event coincides with the
the first collision of the cloud with the CND. The major event
coincides with the second to fourth collisions and with the first crossing
of the streamer with itself. The later and smaller accretion events
coincide with later collisions between the streamer and the CND.
The peak mass accretion rate at the beginning is
$\sim$3\,10$^{-2}$~M$_{\odot}$\,yr$^{-1}$ falling to several
10$^{-4}$~M$_{\odot}$\,yr$^{-1}$ after an elapsed time of 6~Myr.
The peak of the major accretion event is $\sim$10$^{-2}$~M$_{\odot}$\,yr$^{-1}$
decreasing steadily until the end of the simulation.

The evolution of the simulation with 1\% energy loss per collision
is almost identical to that without energy loss. With an energy loss
rate of 10\% per collision the mass is accreted approximately 2 times
faster than for the simulation without energy loss. After the first peak
the mass accretion rate stays at a value $>0.01$~M$_{\odot}$\,yr$^{-1}$.
Then it decreases strongly from 2 to 5~Myr and rises again at 9~Myr.
The collision rate shows the same behaviour. Clouds collide more
frequently during 1 and 2~Myr as for the simulation without energy loss.
At $t=6$~Myr the collision rate begins to rise from 4\,10$^{-4}$~yr$^{-1}$
to 1.5\,10$^{-3}$~yr$^{-1}$ at 9~Myr. This happens, because the cloud
distribution of the infalling streamer symmetrizes and forms a second
circumnuclear disk with a mass of $\sim$1.5\,10$^{4}$~M$_{\odot}$.

\subsubsection{Prograde encounter \label{sec:prograde}}

\begin{figure}
    \resizebox{\hsize}{!}{\includegraphics{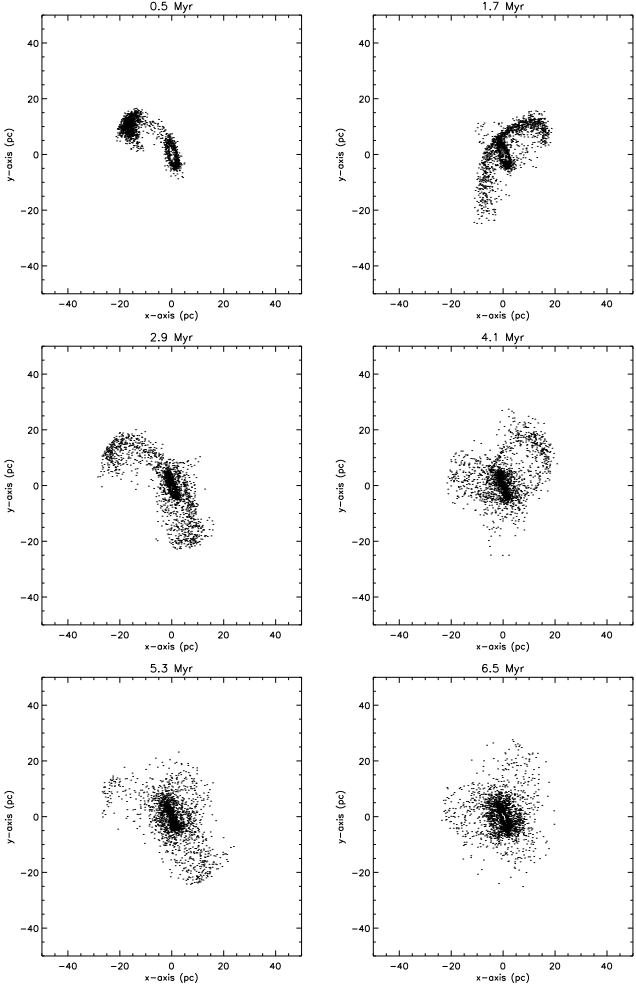}}
      \caption{The evolution of the cloud infall into the
    Galactic Centre as the observer would see it from the Earth.
    The elapsed time is plotted on the top of each frame.
    \label{fig:evolution1}}
\end{figure}

\begin{figure}
    \resizebox{\hsize}{!}{\includegraphics{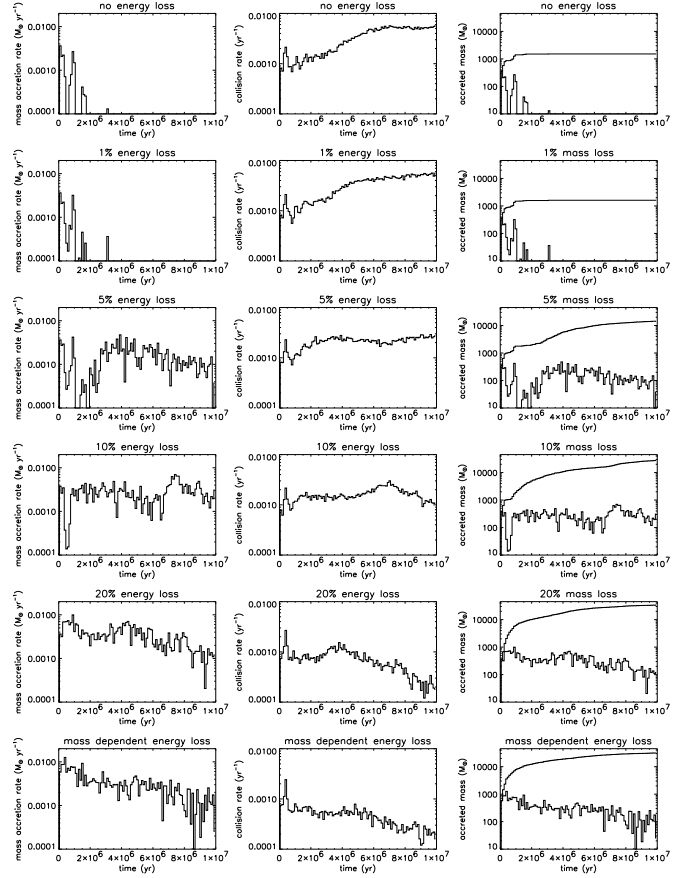}}
      \caption{1st column: mass accretion rate;
    2nd column: number of collisions per timestep $\Delta t$;
    3rd column: accreted mass per timestep $\Delta t$ and accumulate accreted mass.
    Rows: no energy loss, 1\%, 5\%, 10\%, 20\%, mass dependent loss of
    kinetic energy per collision.
    \label{fig:accrete1}}
\end{figure}

The evolution of the system with no loss of kinetic energy per collision
can be seen in Fig.~\ref{fig:evolution1}.
The timestep is $\Delta t$=1.2~Myr.
In difference to the retrograde encounter, the CND is not damaged by
the infalling cloud. During the evolution of the system the cloud--cloud
collisions change the angular momentum of the CND and that of the
infalling cloud. With no energy loss per collision this alignment is
minimum, i.e. at the end of the simulations the majority of the CND clouds
have still their initial angular momentum. We observe at the end
of the simulations a triangular structure with an outer ring--like structure
and the CND inside. Thus, one can still recognize two structure with
different angular momenta. For the simulations with an energy loss
greater than 5\% the situation changes. At the end of the simulation there is
only one kinematical entity, a new CND with an angular momentum between
that of the CND and that of the infalling cloud.
The evolution of the system is shown in Fig.~\ref{fig:accrete1} for 6 different
energy loss rates per collision $\epsilon$: 0\%, 1\%, 5\%, 10\%, 20\%, and a mass
dependent energy loss.
Whereas in the case of a retrograde encounter the general behaviour
of the different quantities were the same for different energy loss rates,
in the case of a prograde encounter these evolutions change dramatically
for different energy loss rates.
In the case of energy loss rates $\epsilon<$1\% the mass accretion rate shows two
peaks of $\dot{M} \sim 2\,10^{-3}$~M$_{\odot}$\,yr$^{-1}$. After a third small peak the
mass accretion rate drops to zero. In the case of $\epsilon$=5\%
there are also two peaks at the beginning followed by an increase between
$t\sim$2~Myr and 4~Myr ($\dot{M} \sim2\,10^{-3}$~M$_{\odot}$\,yr$^{-1}$)
and decreases then slowly. For $\epsilon$=10\%
there is a gap in the mass accretion rate after the first collision between the
infalling cloud and the CND. At later timesteps the mass accretion rate
stays almost constant at a value of $\dot{M} \sim 2\,10^{-3}$~M$_{\odot}$\,yr$^{-1}$.
The gap disappears for $\epsilon$=20\%. After a large first maximum between
0.5 and 1.5~Myr ($\dot{M} \sim 7\,10^{-3}$~M$_{\odot}$\,yr$^{-1}$),
the mass accretion rate has a minimum at $t\sim2.5$~Myr, rises again to
$7\,10^{-3}$~M$_{\odot}$\,yr$^{-1}$ at $t \sim$4~Myr and decreases again.
In the case of a mass dependent energy loss the mass accretion rate
decreases steadily from its initial maximum of $\sim 10^{-2}$~M$_{\odot}$\,yr$^{-1}$
to $\dot{M} \sim 10^{-3}$~M$_{\odot}$\,yr$^{-1}$ after 10~Myr.

\begin{figure}[h]
    \resizebox{\hsize}{!}{\includegraphics{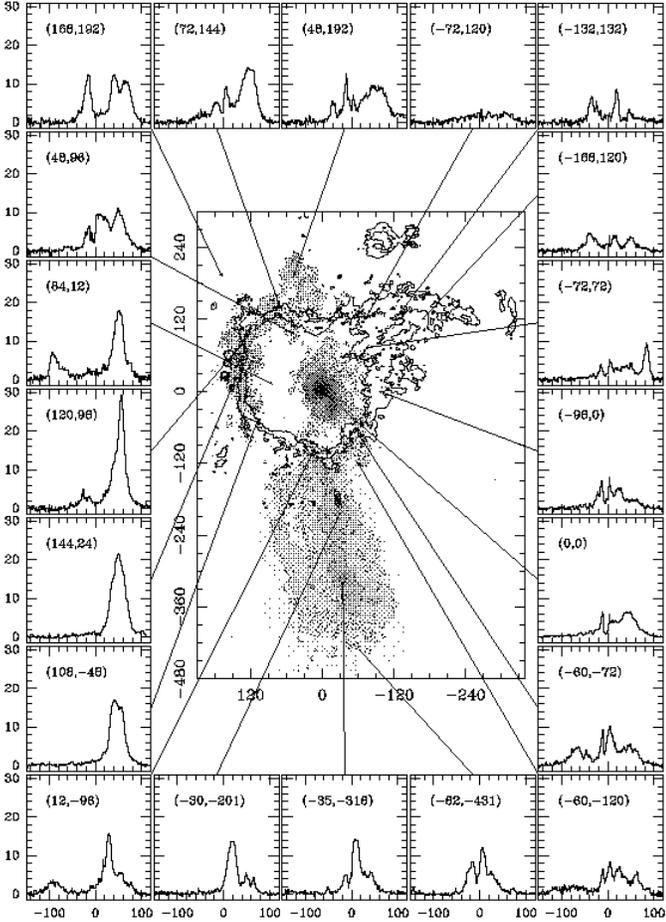}}
      \caption{The Sgr~A Radio and GMC Complex. Greyscale: $\lambda$\,1.3\,mm
with the IRAM MRT (Mezger et al. 1989, Zylka et al. 1990). Contour:
the synchrotron shell Sgr~A East observed at $\lambda$\,6\,cm with the VLA
(Yusef-Zadeh \& Morris, 1987). Frame panels: $^{13}$CO(2-1) IRAM spectra
(Mezger et al. 1996).\label{fig:sgrwestcomplex}}
\end{figure}

\begin{figure}[htb]
    \resizebox{7cm}{7cm}{\includegraphics{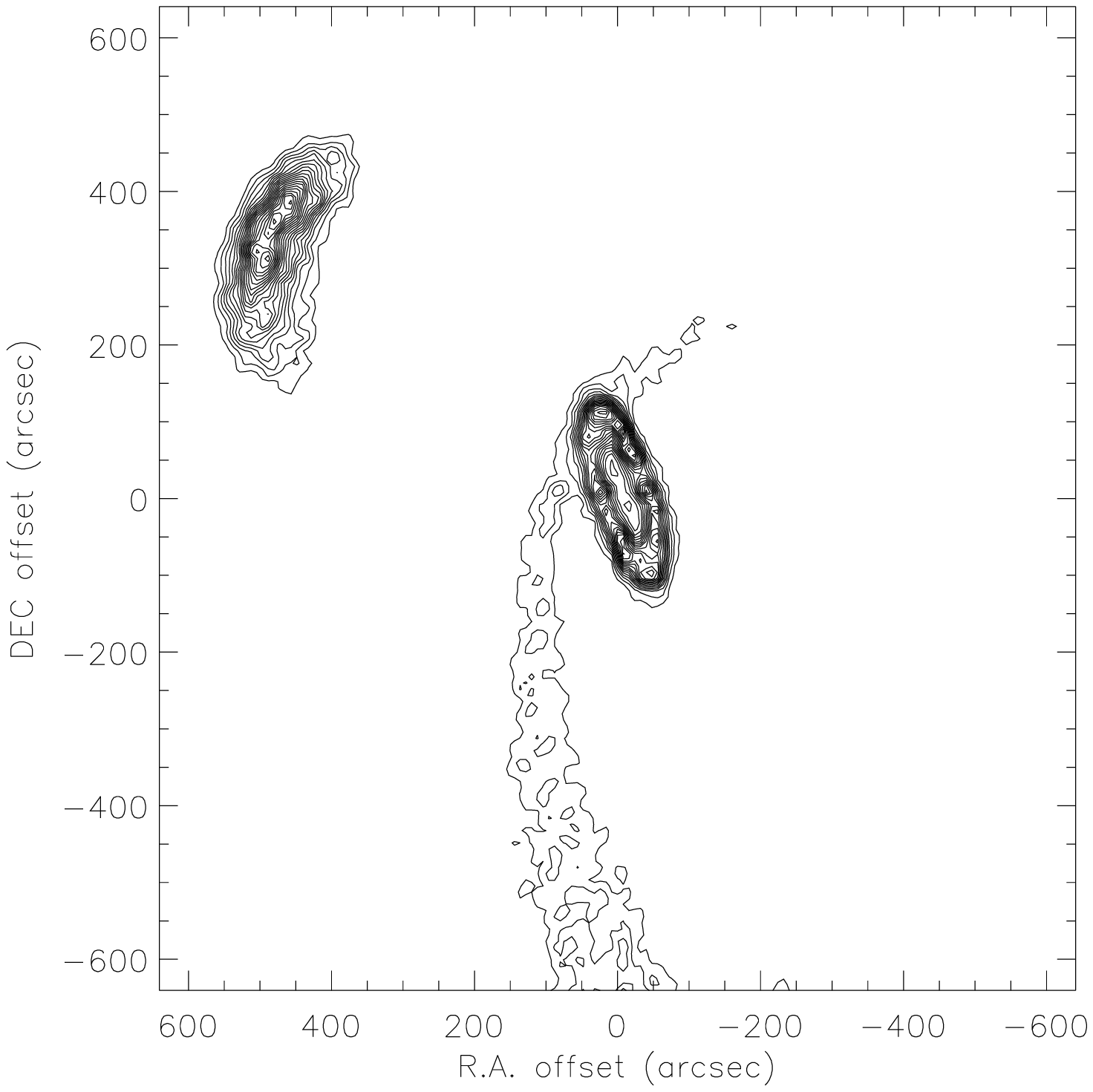}}
    \resizebox{7cm}{7cm}{\includegraphics{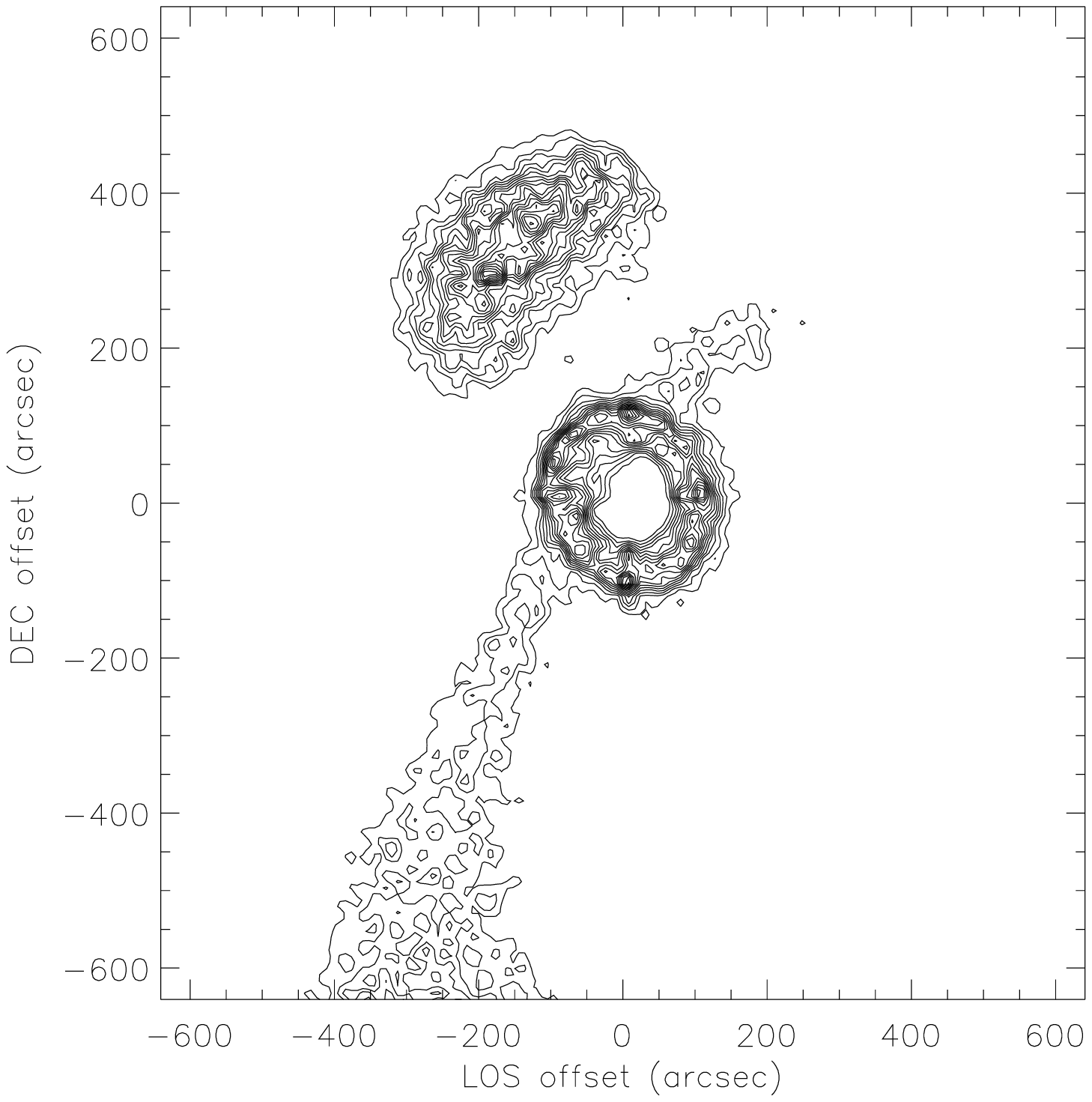}}
      \caption{Left panel: combined snapshots of the high mass simulation
    for the comparison with the Sgr~A cloud complex.
    Right Panel: the same mass distribution but with the line of
    sight (LOS) as the $y$-axis. Smaller distances are nearer to
    the observer.
    \label{fig:snapshot}}
\end{figure}

The evolution of the collision rate also shows big differences for different
energy loss rates $\epsilon$. Whereas the peak ($t_{\rm coll}^{-1} \sim 2\,10^{-3}$~yr$^{-1}$)
due to the first encounter between the infalling cloud and the CND is always there,
the subsequent behaviours are markedly different. For $\epsilon$=0\%
the collision rate begins to rise at $t$=3~Myr until $t$=6~Myr and then stays
constant ($t_{\rm coll}^{-1} \sim 5\,10^{-3}$~yr$^{-1}$).
For $\epsilon$=5\% the collision rate begins to rise at $t \sim$1.5~Myr
and stays constant between 3 and 10~Myr ($t_{\rm coll}^{-1} \sim 2\,10^{-3}$~yr$^{-1}$).
For higher $\epsilon$ the collision rate even decreases at $t \sim 4$~Myr.

\section{Comparison with the Sgr~A cloud complex \label{sec:comarison}}

We have constructed a simple toy model in order to compare the
observed Sgr~A cloud complex (Zylka et al. 1990) to a snapshot of
our simulation (for the observational interpretation see Sect.~\ref{sec:env}).
The Sgr~A Radio and GMC Complex observed at $\lambda$\,1.3\,mm
with the IRAM MRT (Mezger et al. 1989, Zylka et al. 1990), the synchrotron
shell Sgr~A East observed at $\lambda$\,6\,cm with the VLA
(Yusef-Zadeh \& Morris, 1987), and $^{13}$CO(2-1) IRAM spectra
(Mezger et al. 1996) are shown in Fig.~\ref{fig:sgrwestcomplex}.
Since there seems to be at least two different
kinematical features, we have added two snapshots of the tidally
stretched cloud at two different times (0.6~Myr and 1.6~Myr)
together with the CND. This has been done using a simulation with 5 times
more clouds in order to have a particle resolution.
The cloud cross sections were adapted to ensure the same mean collision
time scale as for the simulation with 1500 clouds.
We use a prograde simulation, because it fits best observed line--of--sight
locations of the GMCs in the vicinity of the CND. At these early stages
of evolution, the results do only marginally depend on the energy loss rate
per collision $\epsilon$. We therefore chose $\epsilon$=0.

This composite snapshot can be seen in the left panel of
Fig.~\ref{fig:snapshot}.
It should be stressed here that the aim of this comparison is not
to reproduce each feature in detail but to make a generic picture
of the Sgr~A region. Therefore, it is not troublesome that the
cloud in the east of the Galactic Centre is not as close as
M-0.02-0.07. However, the CND and the streamer in south-north
direction containing M-0.13-0.08 are nicely reproduced.
The mass distribution in direction of the line of sight (LOS)
is shown in the lower panel of Fig.~\ref{fig:snapshot}.
The streamer in south--north direction is located in front of
the Galactic Centre as well as the eastern cloud.

\begin{figure}[htb]
    \resizebox{\hsize}{!}{\includegraphics{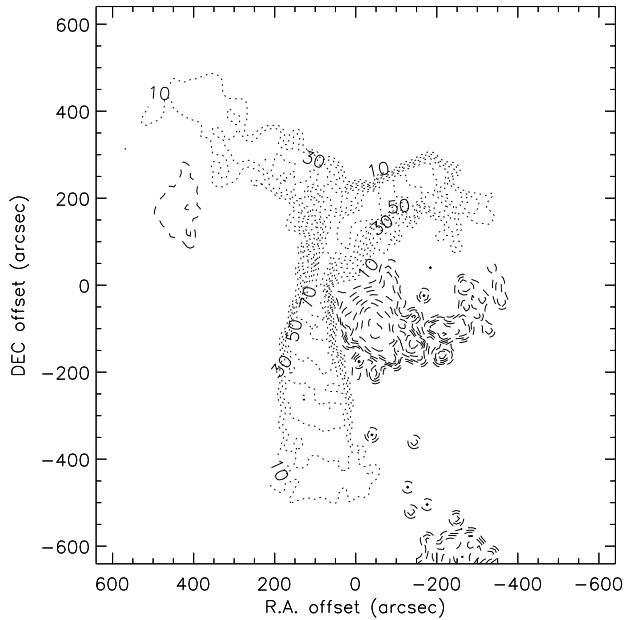}}
      \caption{The radial velocity field of Fig.~\ref{fig:snapshot}.
    Dashed lines represent negative velocities, dotted
    lines positive velocities. The contours are in the range
    of -120--120 km\,s$^{-1}$ with a stepsize of 10 km\,s$^{-1}$.
     \label{fig:snapshotvel}}
\end{figure}
\begin{figure}[h]
    \resizebox{\hsize}{!}{\includegraphics{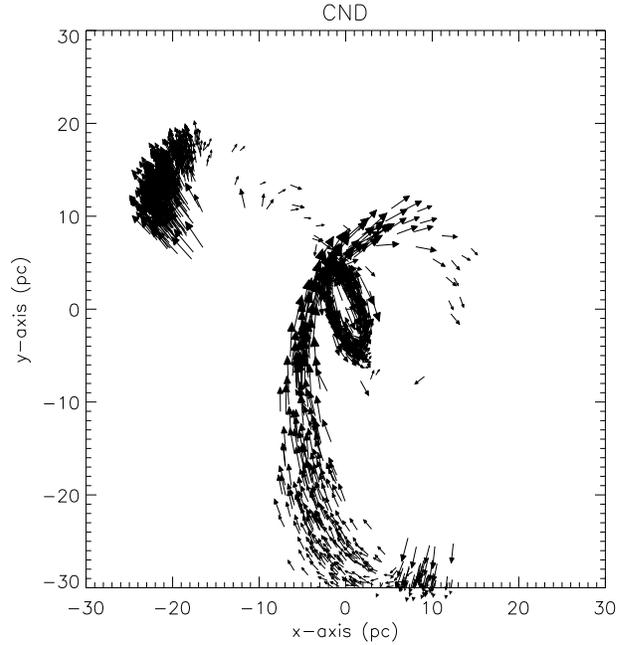}}
      \caption{The velocity field of Fig.~\ref{fig:snapshot}
    in the R.A.--Dec. plane. For clarity only each 10th velocity vector
    is plotted. The CND is rotating counter clockwise.
    Its outer rim is rotating clockwise.
    \label{fig:snapshotmotion}}
\end{figure}

The radial velocity field can be seen in Fig.~\ref{fig:snapshotvel}.
The velocities in the R.A.--Dec. plane are shown in
Fig.~\ref{fig:snapshotmotion}.
The streamer in south--north direction is thus approaching the
Galactic Centre from the front crossing the northern part of the CND.
The eastern cloud is near the apocenter of its elliptical orbit,
the north-western part having already changed the sign of the radial
velocity falling onto the CND from the front.

Zylka et al. (1998) show CS(3-2) spectral line data observed at the
IRAM 30m Telescope covering the central 21$' \times 20'$ with a
spatial resolution of 17$''$. Their galactic longitude--velocity and
galactic latitude--velocity plots can be directly compared with
our simulations. We therefore made plots of exactly the same slices
in $\Delta l$ and $\Delta b$ respectively.

\begin{figure*}[htb]
    \psfig{file=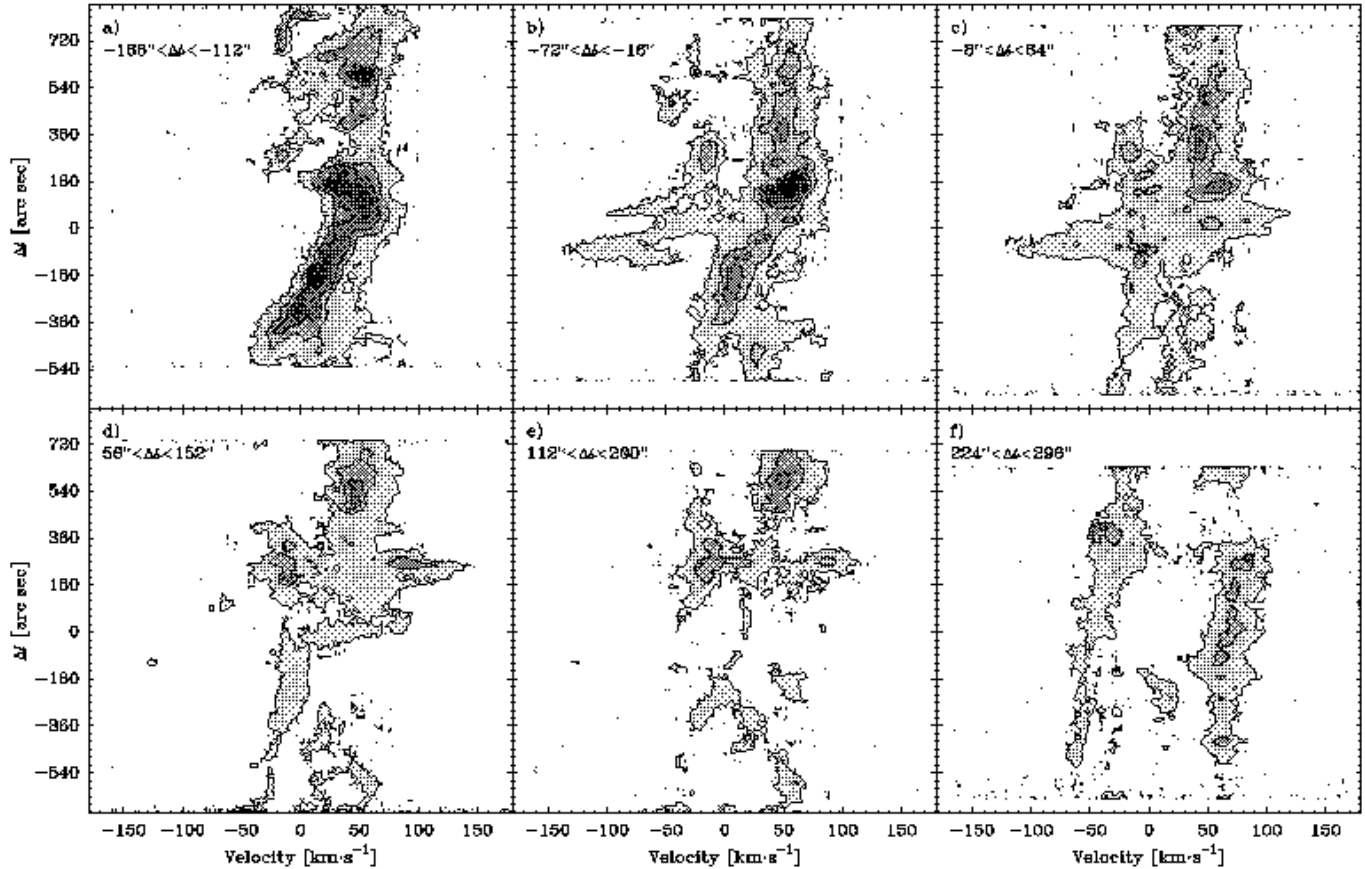,angle=-90,width=\textwidth}
    \caption{Representative galactic longitude--velocity plots ($v$--$\Delta l$)
    of the CS(3--2) emission. The data have been averaged over selected
    ranges in galactic latitudes, $\Delta b$ (Zylka et al. 1998).
    \label{fig:cs3_2avlv_6}}
\end{figure*}

\begin{figure}[htb]
    \resizebox{\hsize}{!}{\includegraphics{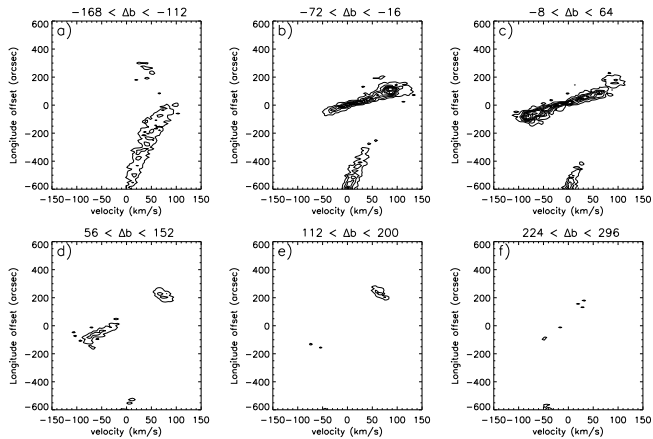}}
      \caption{Corresponding galactic longitude--velocity model plots
    ($v$--$\Delta l$).
    \label{fig:lslice}}
\end{figure}

The galactic longitude--velocity plots of Zylka et al. (1998) are plotted
in Fig.~\ref{fig:cs3_2avlv_6}. The corresponding model diagrams
can be seen in Fig.~\ref{fig:lslice}.
The main features of the CS(3--2) ($v$--$\Delta l$) plots are reproduced by the
model. Obviously, the CS(3--2) intensity does not match the
column density of the model plots. This might be due to the fact that the
CS(3--2) emission depends on the physical condition out of which only
one is the total column density. In Fig.~\ref{fig:lslice}(a) the streamer
in south-north direction is clearly visible. Nevertheless it appears more
prominent in the corresponding CS(3--2) plot (Fig.~\ref{fig:cs3_2avlv_6}(a)).
In Fig.~\ref{fig:lslice}(b) and (c) the CND is prominent.
We miss the observed prominent cloud complex at positive velocities and
positive longitudes almost completely. The eastern model streamer can be
seen in in the upper right part of Fig.~\ref{fig:lslice}(a) and
as an extension of the CND structure at positive velocities in
Fig.~\ref{fig:lslice}(b). Both features have a counterpart in
Fig.~\ref{fig:cs3_2avlv_6}(a) and (b). In contrast, the feature
at negative velocities in Fig.~\ref{fig:lslice}(d), which is due to
the eastern part of the eastern streamer (Fig.~\ref{fig:snapshotvel}),
in Fig.~\ref{fig:lslice}(d) has no counterpart in Fig.~\ref{fig:cs3_2avlv_6}(d).
This implies that the observations might correspond to a slightly
later timestep in the evolution of the eastern model cloud and/or that
its orbital parameters might be slightly different from that of the
observed cloud. Nevertheless, our eastern model cloud represents
a valuable first approach in modelling M-0.02-0.07.

\begin{figure*}[htb]
    \psfig{file=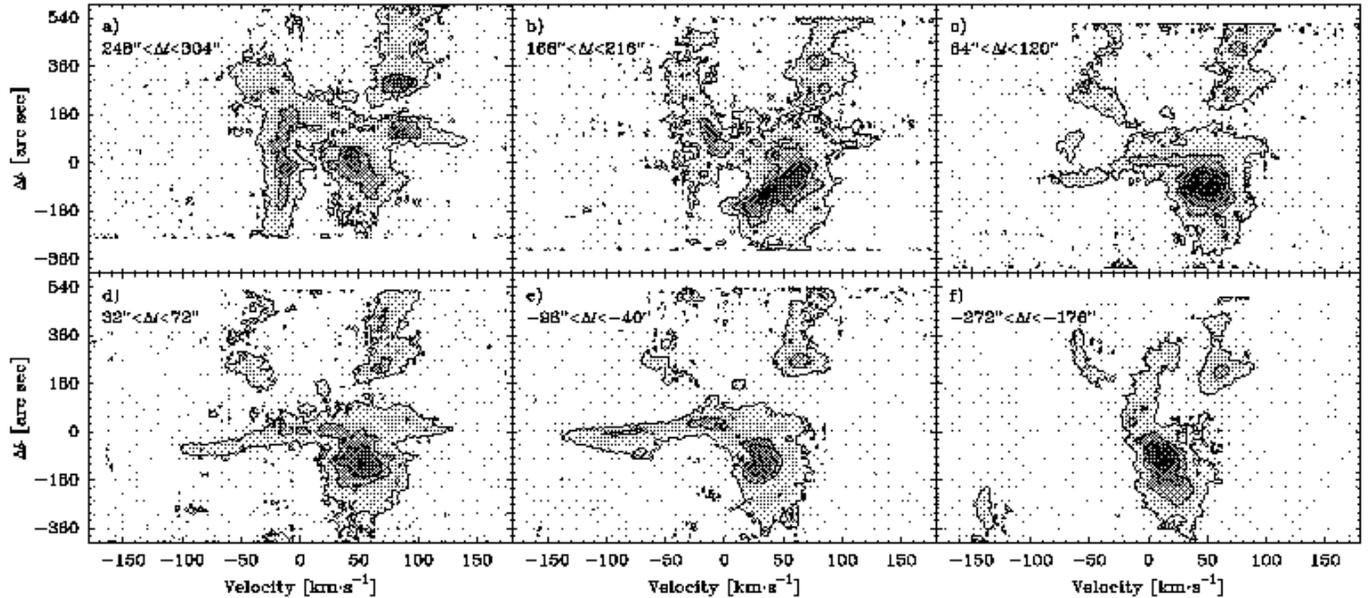,angle=-90,width=\textwidth}
      \caption{Representative galactic latitude--velocity plots ($v$--$\Delta b$)
    of the CS(3--2) emission. The data have been averaged over selected
    ranges in galactic longitude, $\Delta l$ (Zylka et al. 1998).
    \label{fig:cs3_2avbv_6}}
\end{figure*}

\begin{figure}[htb]
    \resizebox{\hsize}{!}{\includegraphics{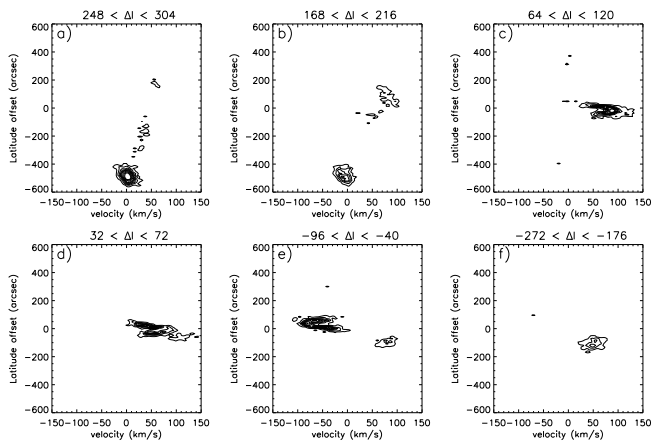}}
      \caption{Corresponding galactic latitude--velocity model plots
    ($v$--$\Delta b$).
    \label{fig:bslice}}
\end{figure}

The galactic latitude--velocity plots of Zylka et al. (1998) are plotted
in Fig.~\ref{fig:cs3_2avbv_6}. The corresponding model diagrams
can be seen in Fig.~\ref{fig:bslice}.

We succeed in reproducing the feature at negative latitudes and positive
velocities (Fig.~\ref{fig:bslice}(e) and (f)), which is caused by the southern
streamer. The model velocities are offset by $\sim$40~km\,s$^{-1}$ from the
observed ones in both diagrams.
The eastern streamer, which can be seen at negative latitudes
in Fig.~\ref{fig:bslice}(a) and (b) has only an observed counterpart
in Fig.~\ref{fig:cs3_2avbv_6}(b). This is due to the fact that
the eastern model streamer is located more to the north than M-0.02-0.07.
The additional CND features at negative latitudes
and negative velocities (Fig.~\ref{fig:bslice}(c) and (d)) might indicate
that the position angle of the CND might differ by some degrees between
the model and the real CND.

\section{Discussion}

In this work we investigated the dynamical behaviour of the
CND in the Galactic Centre. In a first step we treated an isolated
ring--like structure which contains 500 clouds with a cloud mass of
$\sim$30 M$_{\odot}$. Once in equilibrium, the disk shows the
same height with respect to the radius as it is deduced from
mm-observations (G\"{u}sten et al. 1987). We normalized the
number of collisions per unit time with the help of a simple
spherical cloud distribution. The resulting value for the CND
simulation in equilibrium is $t_{\rm coll} \sim 2\,10^{6}$~yr.
Therefore, the isolated CND can be longer lived as it was thought before.
This collision time scale implies a mass accretion rate of
several 10$^{-4}$~M$_{\odot}$\,yr$^{-1}$. We conclude that the current mass
accretion rate at pc scale ranges between 10$^{-3}$ and
10$^{-4}$~M$_{\odot}$\,yr$^{-1}$ (the latter being the value for the isolated CND,
see Vollmer \& Duschl 2001a).

In a second step the environment of the CND was taken into account.
We have constructed a dynamical model where a spherical cloud
falls into the Galactic Centre onto a pre--existing CND.
A retrograde and a prograde encounter were simulated with different
kinetc energy loss rates $\epsilon$ per collisions.
In the case of a retrograde encounter the evolution of the
mass accretion and cloud collision rate during the collision between the cloud
and the CND are not very sensitive to $\epsilon$. We observe a first
maximum during $\Delta t \sim$0.1~Myr followed a larger maximum
($\Delta t >$1~Myr). In the case of a prograde encounter we observe
dramatic changes in the behaviour of these quantities.
For small energy loss rates per collision ($\epsilon \leq$2\%)
only $\sim$2\,10$^{3}$~M$_{\odot}$ are accreted within less than 2~Myr,
whereas the collision rate increases by a factor of
$\sim$5 between 2 and 5~Myr.
For intermediate energy loss rates (5\%$\leq \epsilon \leq$10\%)
we observe an approximately constant mass accretion rate of
$\dot{M} \sim 3\,10^{-3}$~M$_{\odot}$\,yr$^{-1}$ with a gap ($\Delta t \sim$0.5~Myr)
after the first encounter of the cloud with the CND. For $\epsilon$=5\%
there is second broader gap ($\Delta t \sim$2~Myr) $\sim$1~Myr after
the first collision of the cloud with the CND. The collision rate
between the clouds stays constant ($t_{\rm coll}^{-1} \sim 3\,10^{-3}$~yr$^{-1}$)
over the whole evolution.
For $\epsilon \geq$20\% the initial value of the mass accretion rate is
more than a factor of 2 higher than for $\epsilon <$20\% and decreases
steadily during the whole evolution. The collision rate shows a prominent peak
when the cloud collides for the first time with the CND, drops then
to $t_{\rm coll}^{-1} \sim 10^{-3}$~yr$^{-1}$ and varies by a small amount
during the further evolution.

This strong dependence of the mass accretion and the collision rate
on the energy loss rate per collision shows the importance of cooling
mecanisms and the magnetic field, which both influence the energy loss rate.
Whereas the O{\sc i} and C{\sc ii} lines are the major contributors
to radiative energy losses in the shock front during the collision,
the magnetic field retains kinetic energy in the clouds.
Following Krolik \& Begelman (1988) the loss rate of kinetic energy during
a cloud--cloud collision is given by $\epsilon=f\,\xi$, where $f$ is the
fraction of cloud's mass participating in the collision and $\xi$
is the degree of inelasticity. Krolik \& Begelman (1988) estimate
$f=0.2$, which is in good agreement with our findings for the simulations
with a mass dependent energy loss rate. The degree of inelasticity is given by
\begin{equation}
\xi=1-\frac{1}{{\cal M}\sqrt{\beta}-\frac{1}{2}}\ ,
\label{eq:inelastic}
\end{equation}
where $\cal M$ is the Mach number and $\beta$ is the ratio between gas and
magnetic pressure. A Mach number ${\cal M} \sim 20$ and $\beta \sim 1$
gives $\xi = 0.95$. Only for $\beta < 0.1$ more than 20\% of the energy available
in the collision is retained by the magnetic field. The strength of
the magnetic field in the CND is about 1~mG (Yusef-Zadeh et al. 1996),
the cloud densities are of the order 10$^{6}$~cm$^{-3}$, and the gas temperature
within the clouds is $\sim$150~K (Vollmer \& Duschl 2001a).
This leads to $\beta$ of the order of 1. Therefore, we conclude that
magnetic fields might not play a preponderant r\^{o}le for the evolution of
the CND as long as Eq.~(\ref{eq:inelastic}) holds.

Vollmer \& Duschl (2001a) estimate the energy loss rate in the shock
during the collisions to be $\epsilon=$10\%.
Thus, most realistic simulations are those with
$\epsilon$=10--20\% and a mass dependent energy loss rate (lower panels in
Figs.~\ref{fig:accrete} and \ref{fig:accrete1}). For $\epsilon>$10\%
the evolutions of the mass accretion and cloud collision rates of a
retrograde and a prograde encounter are very similar: there is a prominent
peak after the first collision between the cloud and the CND followed by
steady decrease. For $\epsilon=$10\% there are measurable differences
between a retrograde and a prograde encounter (see Sect.~\ref{sec:evolution}).

In a further step we have constructed the observed features of the
Sgr~A cloud complex by adding two snapshots of our simulations.
We can reproduce the observed topography in a satisfying way in
the R.A.--Dec. plane and in galactic longitude/latitude--velocity
planes. This means that the orbit of the recently infalling model cloud is
prograde with respect to the CND and the gas motion in the galactic disk.
The observed warp of the CND (G\"{u}sten et al. 1987) can be naturally explained
by a former prograde encounter (see Sect.~\ref{sec:prograde}).
Based on our model, there is the possibility that one of the streamers has already
passed the Galactic Centre less than 0.5~Myr years ago.
During this former interaction of a cloud and the CND
a small starburst might have taken place at a distance of several pc from
the Galactic Centre. Since the front end of the tidally stretched cloud and the
front of the eastern streamer are
just crossing the CND today, we should be able to find signs of a small
current starburst in the CND. This might be the case for the observed O{\sc i}
cloud (the {\it Tongue}, Jackson et al. 1993). In this case, the incoming
streamer hits the CND at $\sim$3--5~pc. The clouds lose angular
momentum due to collisions and are contracted. The so infalling perturbed
clouds fragment, collapse, and form stars.
These new ionizing O and B stars are still embedded in the part of
the cloud from which they formed. Ott et al. (1999) found
evidence for ongoing star formation in the Central Cavity. Especially,
they identified stars embedded in extended dust clouds within the
Minispiral. The UV radiation of the young hot stellar population
can penetrate efficiently into the fragmented cloud.
Therefore, the gas in this region is predominantly in atomic
form and can not be observed in HCN.

\begin{figure}
    \resizebox{\hsize}{!}{\includegraphics{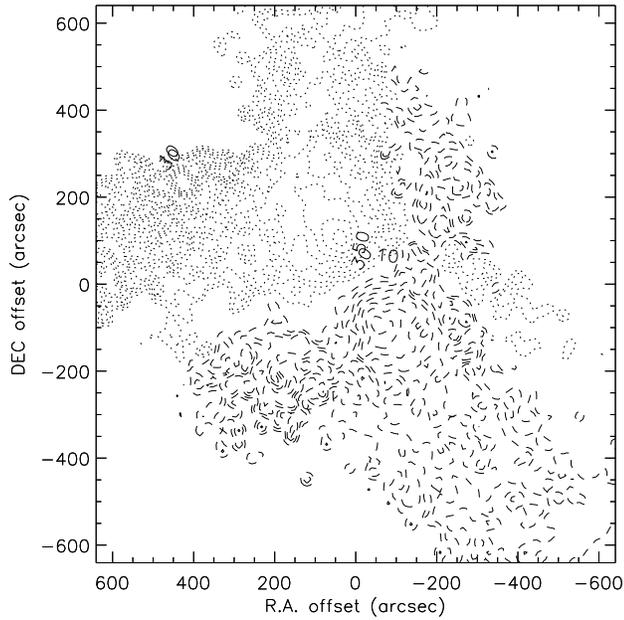}}
      \caption{Velocity field after 10~Myr. This
    configuration can be compared to the velocity field
    of the molecular cloud Sgr~A East Core. The dotted lines represent
    positive radial velocities. The dashed lines represent negative
    radial velocities. \label{fig:sgraeastcore}}
\end{figure}

The remaining question is: how and when did the He{\sc i} star cluster
form? Genzel et al. (1996) found that the He{\sc i} stars are rotating
in a direction opposite to that of Galactic rotation. If the star cluster
was formed through an CND -- molecular cloud interaction and if
the stars' angular momentum reflects that of their parent cloud,
this would imply that the colliding cloud was on a retrograde orbit.
Since we have argued that the energy mass loss rate $\epsilon
\geq$10\% the two last two rows
of Fig.~\ref{fig:accrete} are relevant. If we assume that star formation is
triggered by cloud--cloud collisions, the majority of stars is
formed during the first 2~Myr after the first collision between the cloud
and the CND. Assuming a cloud mass of 30~M$_{\odot}$ this leads to
a total mass of colliding clouds of $\sim$6\,10$^{4}$~M$_{\odot}$ within
2~Myr. On the other hand, if the starburst produces stars with a given IMF,
we can calculate the initial total gas mass needed to form the observed
He{\sc i} stars. Assuming an initial mass function $\Phi=CM^{-2}$ with
cut--offs at 0.2~M$_{\odot}$ and 50~M$_{\odot}$ and normalized
by the observed He{\sc ii} stars:
\begin{equation}
M_{\rm tot}^{\rm HeII}=\int_{10\ {\rm M}_{\odot}}^{50\ {\rm M}_{\odot}} \Phi(m)
{\rm d}M=25 \times M^{\rm HeII} \simeq 500\ {\rm M}_{\odot}
\end{equation}
we deduce total stellar mass produced by the starburst of several
10$^{4}$~M$_{\odot}$. This would imply that 30--50\%
of the colliding gas mass is transformed into stars.
Since the cloud orbit is retrograde with respect to the
rotation of the CND, the colliding clouds lose efficiently
angular momentum and thus fall into the Galactic Centre.
Therefore, we expect that the stars are formed near the inner edge of the CND.
They need $\sim$7~Myr to build a centrally peaked spatial distribution
(Vollmer \& Duschl 2001b). After this time the initial cloud has the appearance
of several streamers (see Fig.~\ref{fig:evolution} at $t$=7~Myr)
and can not be recognized any more as one single cloud.
Nevertheless, we can imagine
that a part of Sgr~A East Core could be identified as this remaining
feature. In order to give an example for the dynamical behaviour of the system
after 10~Myr we plot the velocity field of our simulation after
this time (Fig.~\ref{fig:sgraeastcore}).
The mass distribution and
the velocity field are complicated with one side approaching the
observing and the other one receding. The position and inclination
angle of the orbit of the infalling cloud might not be the one observed.
Since it is difficult to separate really distinct kinematical
features in the observations, the discussed model offers only one plausible
explanation for the complicated kinematical structure of Sgr~A East Core.

Our model still contains great simplifications concerning the
physics of the cloud--cloud collisions (see Sect.~\ref{sec:model}). The effects
of these collisions on the cloud population might change, for example,
with a better knowledge of the detailed radiative magneto-hydrodynamics
which is clearly needed.

The shortcomings concerning the large scale dynamics of our model can in
principle be overcome by modifying the initial conditions for the infalling cloud.
But, since the parameter space for these initial conditions is very
large, it is not the aim of this work to reproduce each feature in great detail
but to show general effects. Nevertheless, we can conclude that the Sgr~A
cloud complex consists of more than one dynamically independent feature.

For the distinction of the dynamically different features in
the Sgr~A cloud complex we followed Zylka et al. (1990)
who made this distinction on observational grounds. It is still possible that
interferences of these features exist. Only an analysis of a complete
data cube of this region together with the model calculation
can shed more light on this question.

\section{Conclusions}

We have made numerical simulations using a collisional N--body code where each
particle has a given mass and radius. These clouds can have inelastic
that are described schematically. We have first normalized the collision
time scale in our code using a spherical configuration.
Simulations of an isolated disk structure corresponding to the
Circumnuclear Disk (CND) in the Galactic Centre yields a mean collision
time scale of one cloud of $t_{\rm coll} \sim 2$~Myr and a mass
accretion rate of $10^{-4}$~M$_{\odot}$\,yr$^{-1} \leq \dot{M} \leq
10^{-3}$~M$_{\odot}$\,yr$^{-1}$.

Since two giant molecular clouds (GMC) are located closer than 50~pc from
the CND, we take in a second step this environment of the CND into account.
We made simulations for a retrograde and a prograde encounter using
different loss rates of kinetic energy during the collisions.
We found that the influence of the energy loss rate on the mass accretion
and cloud collision rates is strongest for a prograde encounter.
We estimate the energy loss rate per collision of the clouds
in the Galactic Centre to be of the order of 10--20\% of the kinetic energy.

The direct comparison of a 1.2~mm maps and CS(3--2) position--velocity diagrams
of a region containing the CND and the GMCs with a model snapshot
shows that the GMCs are on a prograde orbit with respect to the rotation
of the CND. A former prograde encounter of a infalling cloud with
the CND can naturally explain its observed warp.
Since both GMCs might begin to collide with the CND
we expect small current star formation near the inner edge of the CND.

Within our scenario of an encounter between an infalling molecular cloud and a CND,
we point out the possibility that the He{\sc ii} star cluster
has been formed by a retrograde encounter of a cloud with the CND
$\sim$7~Myr ago. The cloud that formed the He{\sc ii} star cluster
has been destroyed by tidal forces and can presently no longer be distinguished as
one single kinematical entity.

\end{document}